\documentclass[aps,floatfix,twocolumn,showpacs,pra]{revtex4}
\usepackage{graphicx,bm,color}
\usepackage[utf8]{inputenc}
\usepackage{amsmath,amsfonts,amssymb}
\usepackage{color}

\newcommand{\be}{\begin{equation}}
\newcommand{\ee}{\end{equation}}
\newcommand{\bea}{\begin{eqnarray}}
\newcommand{\eea}{\end{eqnarray}}

\newcommand{\kk}{\mathbf{k}}
\newcommand{\qq}{\mathbf{q}}

\newcommand{\bb}[1]{\left( #1 \right)}

\newcommand{\dd}{\mathrm{d}}

\begin{document}
\title{Concavity of the collective excitation branch of a Fermi gas in the BEC-BCS crossover}
\author{H. Kurkjian, Y. Castin, A. Sinatra}
\affiliation{Laboratoire Kastler Brossel, ENS, PSL, CNRS, UPMC-Sorbonne Universit\'es and Coll\`ege de France,
Paris, France}

\begin{abstract}
We study the concavity of the dispersion relation $q\mapsto \omega_{\qq}$ of the bosonic excitations of a three-dimensional spin-$1/2$ unpolarized Fermi gas
in the Random Phase Approximation (RPA).
In the limit of small wave numbers $q$ we obtain analytically the spectrum up to order $5$ in $q$. In the neighborhood of $q=0$, a change in concavity between the convex Bose-Einstein condensation (BEC) limit and the 
concave BCS limit takes place at  
$\Delta/\mu\simeq0.869$ [$1/(k_F a)\simeq-0.144$], where $a$ is the scattering length between opposite spin fermions, $k_F$ is the Fermi wave number 
and $\Delta$ the gap according to BCS theory, and $\mu$ is the chemical potential. 
At that point the branch is concave due to a negative fifth order term. 
Our results are supplemented by a numerical study that shows the border between 
the zone of the $(q,\Delta)$ plane where $q\mapsto \omega_{\qq}$ is concave and the zone where it is convex.
\end{abstract}

% 03.75.Kk	Dynamic properties of condensates; collective and hydrodynamic excitations, superfluid flow
% 67.85.Lm	Degenerate Fermi gases
% 47.37.+q	Hydrodynamic aspects of superfluidity; quantum fluids
\maketitle

%%%%%%%%%%%%%%%%%%%%%%%%%%%%%%%%%%

\section{Introduction}

%%%%%%%%%%%%%%%%%%%%%%%%%%%%%%%%%%

Cold atomic gases offer a broad flexibility of the microscopic parameters in exploring the many-body problem. 
In particular, in spin-$1/2$ Fermi gases, the interaction strength can be adjusted experimentally using Feshbach resonances without
inducing strong three-body losses.
This degree of freedom, unique among Fermi systems, allowed cold-atom experiments \cite{Thomas2002,Salomon2003,Grimm2004,Grimm2004b,Ketterle2004,Ketterle2005,Salomon2010,Zwierlein2012,Grimm2013} 
to study the crossover between a superfluid of 
Cooper pairs in the so-called Bardeen-Cooper-Schrieffer (BCS) regime of interaction and a superfluid of tightly bound, almost bosonic, dimers 
in the Bose-Einstein Condensation (BEC) regime. 
Another advantage of cold atomic gases is the simple theoretical description of the interactions that the cold and dilute regime in which they occur allows for.
For a Fermi gas in two spin states $\uparrow$ and $\downarrow$, one can show that the only significant interactions at low temperature and weak density 
occur between opposite spins fermions in the $s$ wave and can be fully characterized by a single parameter called the scattering length and denoted by $a$.
In this propitious theoretical framework, entirely analytical studies of experimentally accessible properties of the gas are possible. This article is one of those.

At zero temperature, the three-dimensional spatially homogeneous unpolarized Fermi gas is fully paired and
its excitation spectrum consists of two branches: a fermionic branch of excitation of the 
internal degrees of freedom of the $\uparrow\downarrow$ pairs of fermions 
and a bosonic branch of excitation of their center-of-mass motion, which has a phononic behavior in the long wavelength limit.
The latter branch is sometimes said to be collective since it involves a large number of the fermionic modes of internal excitations of the pairs. 
The fermionic branch is described to lowest order by the BCS theory. To tackle the bosonic branch several approaches have been proposed: 
The Random Phase Approximation (RPA) of Anderson \cite{Anderson1958}, 
a Gaussian approximation of the action in a path integral framework \cite{Strinati1998,Randeria2014}, 
a Green's function approach associated with a diagrammatic approximation \cite{CKS2006}, and a linearization of the time dependent BCS variational equations \cite{KCS2015}.
Remarkably, those theories all lead to the same approximate spectrum of bosonic excitations, which they describe by the same implicit equation.

The concavity of this spectrum has been studied in the weak-coupling BCS limit \cite{Strinati1998} and, in a qualitative way, in the rest of the BEC-BCS crossover \cite{CKS2006,Castin2015,Salasnich2015}. 
A complete quantitative study is thus missing, a gap that this article intends to bridge.
In particular, we obtain the spectrum analytically up to order $5$ in the wave number $q$ of the center of mass of the pairs. This allows us to conclude on the concavity of the branch of excitation
in a neighborhood of $q=0$ over the whole BEC-BCS crossover. 

Several physically relevant problems can be addressed after our study. First, the processes that dominate the collective mode damping at low temperature can be identified. 
If the branch is convex over a neighborhood of $q=0$ then the Landau-Beliaev $2\,\mbox{phonons} \leftrightarrow 1\,\mbox{phonon}$ interaction processes \cite{Beliaev1958,Shlyapnikov1998} dominate 
while if it is concave, those processes are forbidden by momentum and energy conservation and the Landau-Khalatnikov $2\,\mbox{phonons}\leftrightarrow2\,\mbox{phonons}$ 
processes \cite{Khalatnikov1949} take over. 
At low temperature the contribution of the gapped fermionic branch to the collective mode damping is exponentially small  \cite{Liu2011}. 
Second, the quantitative knowledge of the concavity parameter $\gamma$ is required in order to predict the phonon damping rate 
due to the $2\,\mbox{phonons}\leftrightarrow1\,\mbox{phonon}$
processes in the convex case beyond the quantum hydrodynamics approximation \cite{Salasnich2015}, or due to the $2\,\mbox{phonons}\leftrightarrow2\,\mbox{phonons}$ processes in the concave case where the effective 
interaction predicted by quantum hydrodynamics involves virtual $2\,\mbox{phonons}\leftrightarrow1\,\mbox{phonon}$ processes and depends on $\gamma$.
Finally, the knowledge of $\gamma$ gives access to the phase diffusion coefficient of the condensate of pairs, a quantity responsible for an intrinsic and fundamental limit to the coherence time
of the gas \cite{KCS2015}.

%%%%%%%%%%%%%%%%%%%%%%%%%%%%%%%%%%

\section{The RPA equation on the excitation spectrum}

%%%%%%%%%%%%%%%%%%%%%%%%%%%%%%%%%%

The RPA equation yielding implicitly the energy $\hbar \omega_\qq$ of the collective excitations as a function of their wave vector $\qq$ is 
\be
I_{++}(\omega_\qq,q) I_{--}(\omega_\qq,q) = {\hbar^2\omega^2_\qq} \left[I_{+-}(\omega_\qq,q)\right]^2.
\label{eq:dedispersion}
\ee
The collective nature of the bosonic modes is visible in the quantities $I_{\sigma\sigma'}$
which are integrals on the relative wave vector $\kk$ of the pairs and depend on $\omega_\qq^2$ {\sl via} the denominator of their integrand \cite{CKS2006}: 
\begin{widetext}
\bea
I_{++}(\omega,q)&=&\int \dd^3k\left[\frac{(\epsilon_{\kk+\qq/2}+\epsilon_{\kk-\qq/2})(U_{\kk+\qq/2}U_{\kk-\qq/2}+V_{\kk+\qq/2}V_{\kk-\qq/2})^2}{\hbar^2\omega^2
-(\epsilon_{\kk+\qq/2}+\epsilon_{\kk-\qq/2})^2}+\frac{1}{2\epsilon_\kk}\right], \\
I_{--}(\omega,q) &=&\int \dd^3k\left[\frac{(\epsilon_{\kk+\qq/2}+\epsilon_{\kk-\qq/2})(U_{\kk+\qq/2}U_{\kk-\qq/2}-V_{\kk+\qq/2}V_{\kk-\qq/2})^2}
{\hbar^2\omega^2-(\epsilon_{\kk+\qq/2}+\epsilon_{\kk-\qq/2})^2}+\frac{1}{2\epsilon_\kk}\right], \\
I_{+-}(\omega,q)&=&\int \dd^3k\frac{(U_{\kk+\qq/2}U_{\kk-\qq/2}+V_{\kk+\qq/2}V_{\kk-\qq/2})(U_{\kk+\qq/2}U_{\kk-\qq/2}-V_{\kk+\qq/2}V_{\kk-\qq/2})}{\hbar^2\omega^2-(\epsilon_{\kk+\qq/2}+\epsilon_{\kk-\qq/2})^2}.
\eea
\end{widetext}
We have introduced the amplitudes $U_\kk$ and $V_\kk$ and the eigenenergies $\epsilon_\kk$ of the fermionic eigenmodes of the BCS theory \cite{BCS1957}:
\bea
\epsilon_\kk &=& \sqrt{\bb{\frac{\hbar^2k^2}{2m}-\mu}^2+\Delta^2}, \\
U_{\kk} &=& \sqrt{\frac{1}{2}\bb{1+\frac{\frac{\hbar^2k^2}{2m}-\mu}{\epsilon_\kk}}}, \\
V_{\kk} &=& \sqrt{\frac{1}{2}\bb{1-\frac{\frac{\hbar^2k^2}{2m}-\mu}{\epsilon_\kk}}},
\eea
where $m$ is the mass of a fermion. The two natural parameters of the BCS theory, with which we express the energy $\hbar\omega_\qq$, are the chemical potential $\mu$, identical for the two spin states,
and $\Delta$, the gap in the BCS spectrum of the fermionic excitations when $\mu$ is positive.
If needed, they can be replaced by the scattering length $a$ and the total density $\rho$ of the gas by inverting the two following relations \cite{Varenna,KCS2013}:
\bea
 \frac{m}{4 \pi \hbar^2 a}  &=& \int \frac{d^3k}{(2\pi)^3} \bb{ \frac{m}{\hbar^2 k^2}  -\frac{1}{2\epsilon_\kk}}, \label{eq:g0} \\
 \rho                                     &=&  \int \frac{d^3k}{(2\pi)^3} 2|V_\kk|^2. \label{eq:rho}
\eea
In practice, the Fermi wave number $k_F$ defined by $\rho = k_F^3/(6\pi^2)$
is often used instead of the density.
%%%%%%%%%%%%%%%%%%%%%%%%%%%%%%%%%%

\section{Global numerical study of the concavity}

%%%%%%%%%%%%%%%%%%%%%%%%%%%%%%%%%%

\begin{figure}[htb]
\includegraphics[width=0.4\textwidth,clip=]{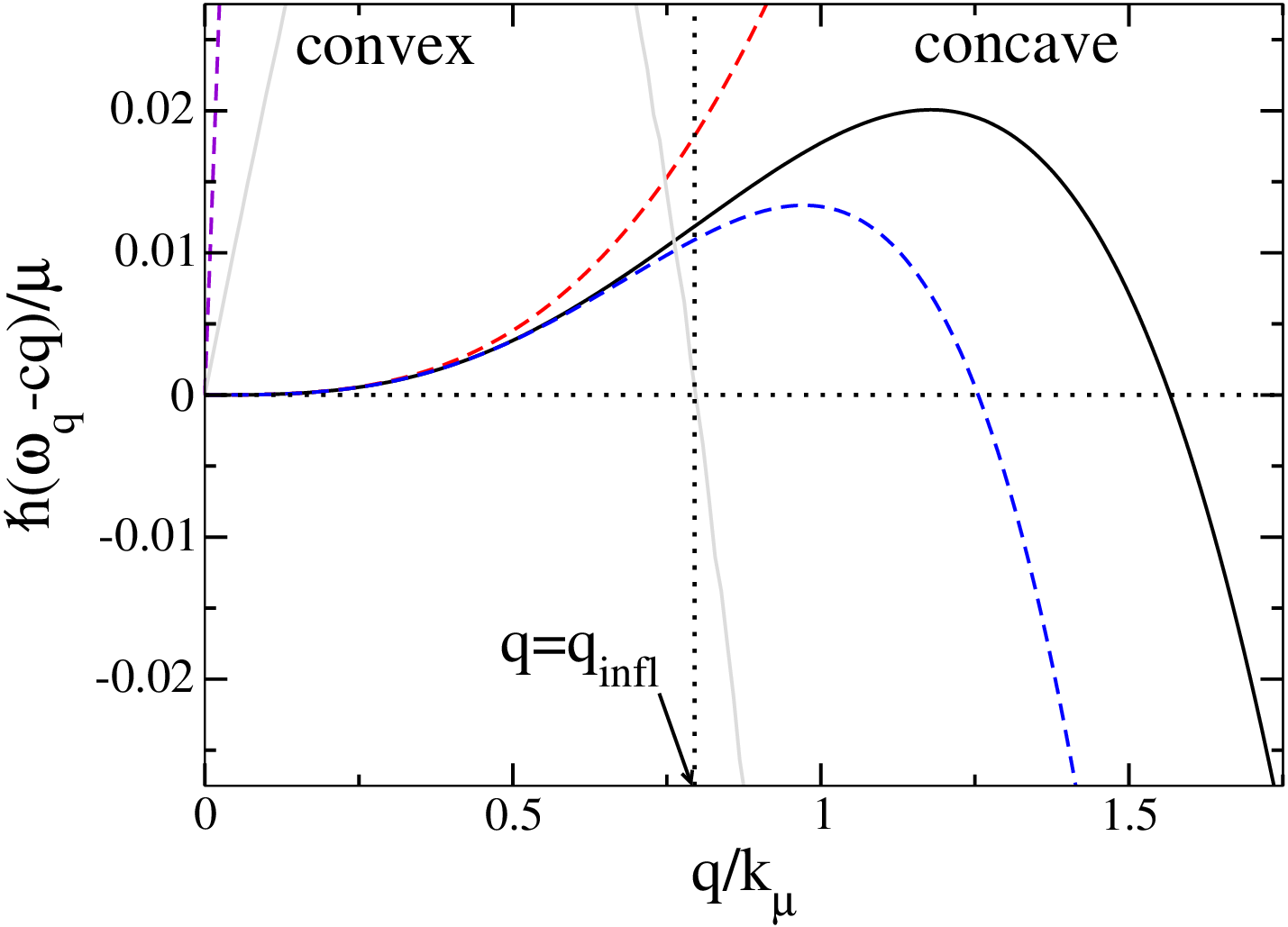}
\caption{(Color online) 
At unitarity $(k_F a)^{-1}=0$, which is here $\Delta/\mu=1.162$, the dispersion relation of the collective excitations 
is plotted as a black solid line {\sl after subtraction} of its phononic part $cq$ to better reveal its concavity properties.
The wave number $q$ is in units of $k_\mu=(2m\mu)^{1/2}/\hbar$.
The linear part $q\mapsto cq$ is shown for comparison
(steep violet dashed straight line), as well as the cubic (red dashed upper line) and the quintic (blue dashed lower line) approximations
of $\omega_\qq-cq$, see Eq.~(\ref{eq:fleur}). The second order derivative $\dd^2\omega_{\qq}/\dd q^2$, plotted as a gray solid line, vanishes
at the inflection point $q_{\rm infl}\simeq 0.795 k_\mu$ marked by the vertical dotted line. To the left (right) of this point 
the dispersion relation $q\mapsto \omega_\qq$ is convex (concave).
\label{fig:disp}}
\end{figure}

From a numerical solution of the dispersion equation \eqref{eq:dedispersion} we obtain the dispersion relation
$q\mapsto \omega_\qq$ over its existence domain. We show an example in Figure \ref{fig:disp} in the unitary limit $1/(k_F a)=0$,
where $\Delta/\mu\simeq 1.162$ according to the BCS theory. Rather than $q\mapsto \omega_\qq$ itself, we plot as a black solid line
the function $q\mapsto\omega_\qq-cq$, where $c$ is the speed of sound and $q\mapsto cq$ is the linear part of the spectrum.
The concavity properties of this function are the same as those of $q\mapsto\omega_\qq$, but they are more visible
graphically at low $q$, because they are not masked by the dominant linear part, which anyway plays no role in the selection
of the damping processes mentioned at the end of the Introduction. In the figure, it is apparent that the dispersion relation
exhibits an inflection point at $q/k_\mu\simeq 0.795$, where $k_\mu=(2m\mu)^{1/2}/\hbar$, separating a low-$q$ interval over which
the dispersion relation is convex and a high-$q$ interval over which it is concave.
As a consequence, according to the RPA, 
the leading damping processes of the collective excitations of the unitary gas at low temperature are the
Beliaev-Landau $2\,\mbox{phonons} \leftrightarrow 1\,\mbox{phonon}$ processes.

\begin{figure}[htb]
\includegraphics[width=0.49\textwidth,clip=]{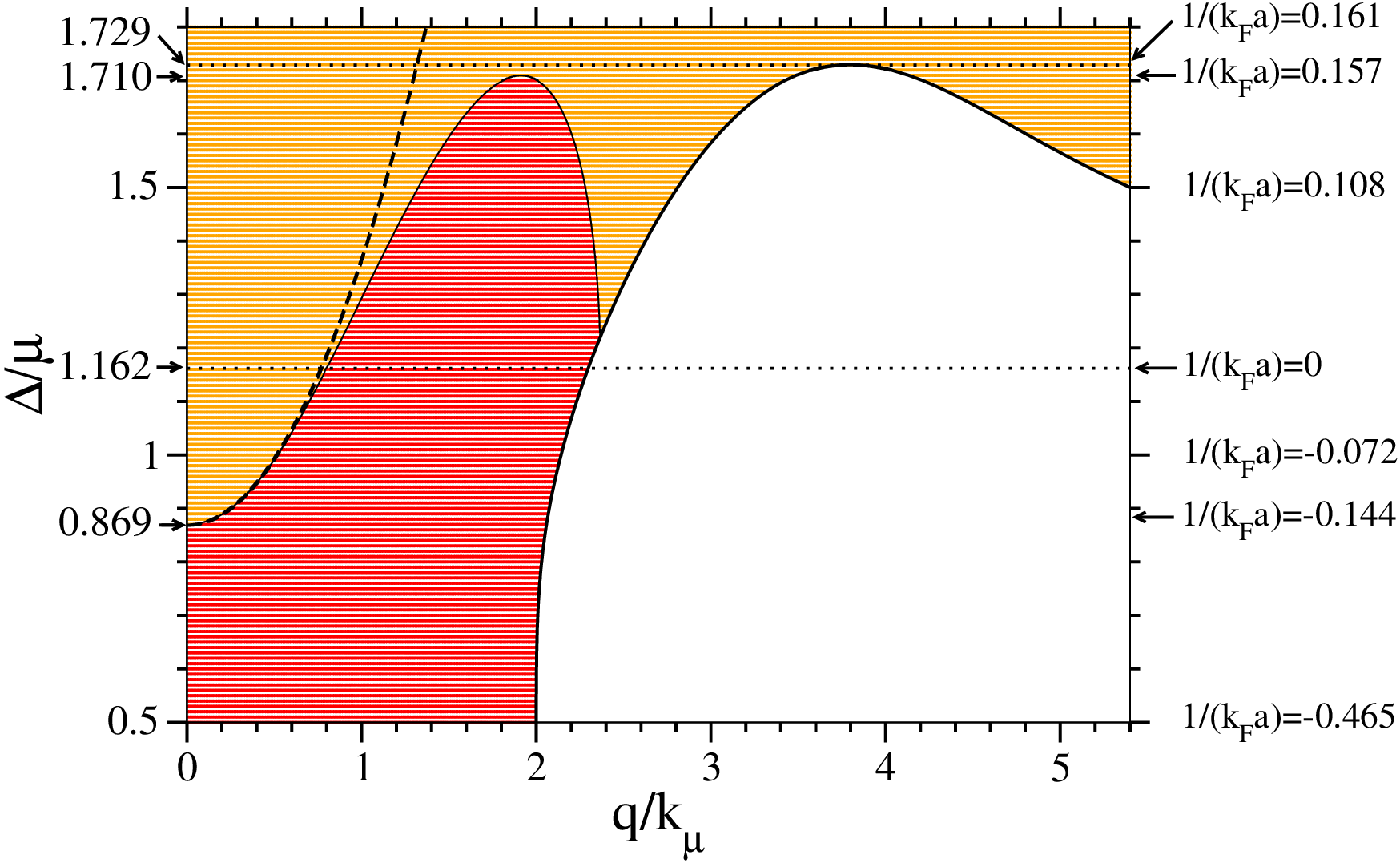}
\caption{(Color online) 
Local concavity of the bosonic branch $q\mapsto \omega_\qq$ depending on $\Delta/\mu$ and on the wave number $q$ in units of $k_\mu=(2m\mu)^{1/2}/\hbar$.
The values of $1/(k_F a)$ corresponding to those of $\Delta/\mu$ are given on the right vertical axis. Red points show that the branch is locally concave and
orange points that the branch is locally convex. 
The thick solid line is the border of the existence domain of the solutions of the dispersion equation \eqref{eq:dedispersion}.
The black dashed line is the low-$q$ analytical prediction (\ref{eq:qinfl}) of the boundary between the red zone and the orange zone, that is of the locus of the inflection points 
of the curve $q\mapsto\omega_\qq$, indicated by a thin solid line. The ordinate $\Delta/\mu=0.869$ of the point where this line meets the $q=0$ axis, 
and the ordinate $\Delta/\mu=1.710$ above which the concavity zone disappears are indicated by arrows. For the values of $\Delta/\mu$
or $1/(k_F a)$ in between the two dotted lines indicated by arrows, 
the $q$-existence domain of the solution of equation \eqref{eq:dedispersion} is not simply connected.
The lower dotted line indicates the unitary limit, where the scattering length diverges $|a|\to+\infty$. \label{fig:conc}}
\end{figure}

Figure \ref{fig:conc} synthesizes our numerical results on the concavity of the bosonic branch for all values of $\Delta/\mu$
and $q/k_\mu$.
Several domains can be identified depending on the values of $\Delta/\mu$, or equivalently $1/(k_F a)$. (i) When the scattering length is negative $a<0$, 
that is for $\Delta/\mu<1.162$ (lower dotted line), the existence domain of the solution to Eq.\eqref{eq:dedispersion} as a function of $q$ is compact 
and simply connected \cite{CKS2006},
hence of the form $[0,q_{\rm sup}]$. 
The dispersion relation is entirely concave for $\Delta/\mu<0.869$ while for $0.869<\Delta/\mu<1.162$ it is first convex at small $q$ and then concave. Between those two zones 
it goes through an inflection point, whose position $q_{\rm infl}(\Delta/\mu)$ we compute analytically in the small $q$ limit 
[see the black dashed line and Eq.(\ref{eq:qinfl}) of section \ref{sec:analytique}].  
(ii) On the other side of the resonance ($a>0$) and up to $\Delta/\mu=1.729$ (upper dotted line), the existence domain of the 
solution to Eq.\eqref{eq:dedispersion} splits up into two connected components $[0,q_{\rm sup}]$ and $[q_{\rm inf},+\infty[$ \cite{CKS2006}. 
While the branch is always convex in the second component $[q_{\rm inf},+\infty[$, the first one $[0,q_{\rm sup}]$ exhibits interesting variations:
From small $q$ to high $q$, the branch is convex and then concave for $1.162<\Delta/\mu<1.22$, convex, concave and then convex again
for $1.22<\Delta/\mu<1.710$ and finally entirely convex for $1.710<\Delta/\mu<1.729$. 
(iii) When $\Delta/\mu$ is greater than $1.729$, or negative, the two components of the existence domain merge
and a solution to Eq.\eqref{eq:dedispersion} exists for all $q$ \cite{CKS2006}.
The branch is then entirely convex.

All these numerical values are predicted by the RPA or the BCS theory. They are therefore approximate. Up to now the only value that can be compared to experiments is 
that of $\Delta/\mu$ at the unitary limit: From the measured values $\Delta\simeq 0.44 \hbar^2 k_F^2/(2m)$ \cite{Ketterle2008} and
$\mu\simeq 0.376 \hbar^2 k_F^2/(2m)$  \cite{Zwierlein2012}
we get $\Delta/\mu\simeq 1.17$, which is remarkably close to the BCS-theory prediction $\Delta/\mu\simeq 1.162$.
One must also keep in mind that the RPA spectrum results from a linearized treatment of the pair-field quantum fluctuations, which neglects the interactions
among the bosonic quasiparticles. In reality, these interactions will shift the eigenenergies $\hbar\omega_\qq$. They will also give rise, even at zero temperature,
to an imaginary part in $\omega_\qq$, corresponding to a finite lifetime of the excitations, provided that the concavity of the dispersion relation
allows for resonant $1\,\mbox{phonon}\to 2\, \mbox{phonons}$ Beliaev processes \cite{Salasnich2015}.

%%%%%%%%%%%%%%%%%%%%%%%%%%%%%%%%%%

\section{Analytical study of the concavity in the long wavelength limit}
\label{sec:analytique}
%%%%%%%%%%%%%%%%%%%%%%%%%%%%%%%%%%
The dispersion relation can be obtained analytically in the long wavelength limit  $q\rightarrow0$. To this end we expand the eigenenergy of the collective mode
up to order $5$ in $q$:
\be
\hbar \omega_\qq \underset{q\to 0}{=} \hbar c q \left[1+\frac{\gamma}{8} \bb{\frac{\hbar q}{mc}}^2+\frac{\eta}{16} \bb{\frac{\hbar q}{mc}}^4 + O\bb{\frac{\hbar q}{mc}}^6
\right]. \label{eq:fleur}
\ee
To lowest order, as for any superfluid system, the energy is phononic with a sound velocity given by the hydrodynamic expression
\be
mc^2=\rho \bb{\frac{\partial\mu}{\partial\rho}}_a,\label{eq:vitesseson}
\ee
where the derivative is taken for a fixed scattering length $a$, as indicated by the notation. When applied to the approximate equation of state \eqref{eq:rho},
the hydrodynamic expression \eqref{eq:vitesseson} gives the RPA sound velocity, as shown in reference \cite{CKS2006} by an expansion of the solution $\omega_\qq$ 
of equation (\ref{eq:fleur}) up to first order in $q$. 
We give here an explicit expression in the form of a rational fraction
\be
\frac{mc^2}{\mu} = \frac{2(xy+1)}{3(y^2+1)},\label{eq:vitesseson2}
\ee
in terms of the variables 
\be
x = \frac{\Delta}{\mu} \ \ \ \ \mbox{and}\ \ \ \ 
y = \bb{\frac{\partial\Delta}{\partial\mu}}_a.
\ee
In turn, the $y$ variable is written as a function of $x$,
\be
y =\frac{\int \dd^3 k \frac{\hbar^2k^2/(2m)-\mu}{\epsilon_\kk^3}}{\int \dd^3 k\frac{\Delta}{\epsilon_\kk^3}} =\frac{\int_0^{+\infty} \dd u\frac{u^2(u^2-x^{-1})}{[(u^2-x^{-1})^2+1]^{3/2}}}{\int_0^{+\infty} \dd u \frac{u^2}{[(u^2-x^{-1})^2+1]^{3/2}}}
\label{eq:dDeltadmu}
\ee
by taking the derivative of equation \eqref{eq:g0} with respect to $\mu$ at fixed scattering length $a$ and by expressing the wave vectors in 
units of $k_\Delta=(2m\Delta)^{1/2}/\hbar$ to form the dimensionless integration variable $u$.
The integrals over $u$ in the numerator and in the denominator of the right-hand side of equation \eqref{eq:dDeltadmu} may be expressed in terms of complete
elliptic integrals of the first and second  kinds \cite{Strinati1998}.
At the unitary limit one has $y=x$, since $\Delta$ and $\mu$ are proportional due to scale invariance.

To obtain the expression \eqref{eq:vitesseson2} of the reduced sound velocity, we take the derivative of the equation of state \eqref{eq:rho} with respect to $\mu$
at fixed $a$ and we express all resulting integrals as functions of $x$ and $y$ using \eqref{eq:dDeltadmu} and the relation
\be
\frac{\int \frac{\dd^3 k}{(2\pi)^3} \frac{\Delta^3}{\epsilon_\kk^3}}{\rho}=\frac{\int_0^{+\infty} \dd u\frac{u^2}{[(u^2-x^{-1})^2+1]^{3/2}}}{\int_0^{+\infty} \dd u\, u^2\bb{1-\frac{u^2-x^{-1}}{[(u^2-x^{-1})^2+1]^{1/2}}}}
=\frac{3x/2}{1+xy},
\label{eq:thetasurN}
\ee
which may be derived by integrating by parts the integral over $u$ in the denominator ($u^2$ is the function to be integrated).

To obtain the dimensionless coefficients $\gamma$ and $\eta$ of the terms of $\hbar\omega_\qq$ of higher order in $q$ in equation \eqref{eq:fleur},
we cannot rely on known thermodynamic expressions and we must face the laborious expansion of equation (\ref{eq:dedispersion}) in powers of $q$.
Still the result can be put into a rational form in terms of $x$ and $y$:
\bea
\gamma           &=& \frac{\sum_{i=0}^4 P_i(y) x^i}{135 x^2 \left(x^2+1\right) \left(y^2+1\right)^3}, \label{eq:gamma} \\
\eta             &=& \frac{\sum_{i=0}^8 Q_i(y) x^i}{1020600 \left(y^2+1\right)^6 x^4\left(x^2+1\right)^2}. \label{eq:eta}
\eea
The $P_i(y)$ and $Q_i(y)$ polynomials that appear as coefficients of $x^i$ in the numerators of $\gamma$ and $\eta$ are given by
\bea
P_0(y) &=& -4 \left(13 y^4+16 y^2+8\right), \notag \\
P_1(y) &=& 4 y \left(13 y^4+41 y^2+8\right), \notag \\
P_2(y) &=& 50 y^6-21 y^4-252 y^2-61,\\
P_3(y) &=& 2 y \left(y^4+32 y^2+71\right), \notag \\
P_4(y) &=& 35 y^6+56 y^4-13 y^2-54, \notag 
\eea
\begin{widetext}
and by
\bea
Q_0(y) &=& 16 \left(7745 y^8+19528 y^6+20304 y^4+8384 y^2+1088\right), \notag \\
Q_1(y) &=& 32 y \left(2857 y^8+67 y^6-3186 y^4-7920 y^2-2624\right), \notag \\
Q_2(y) &=& -8 \left(12882 y^{10}+28061 y^8-26936 y^6+7221 y^4-24496 y^2-5232\right), \notag \\
Q_3(y) &=& -8 y \left(8456 y^{10}-9859 y^8+9977 y^6+145295 y^4+3523 y^2+23720\right), \notag \\
Q_4(y) &=& -17500 y^{12}-247996 y^{10}-1249743 y^8-1341332 y^6+337202 y^4-694392 y^2+18321, \\
Q_5(y) &=& -4 y \left(25564 y^{10}+36027 y^8-66984 y^6+92206 y^4+387932 y^2-56121\right), \notag \\
Q_6(y) &=& -2 \left(12250 y^{12}+115637 y^{10}+558246 y^8+1071518 y^6+589478 y^4-248499 y^2+53082\right), \notag \\
Q_7(y) &=& -4 y \left(12957 y^{10}+33764 y^8-41904 y^6-173106 y^4-96189 y^2+53406\right), \notag \\
Q_8(y) &=& -8575 y^{12}-44544 y^{10}-149742 y^8-360644 y^6-477615 y^4-270756 y^2-20412. \notag 
\eea
\end{widetext}
Our analytical expressions \eqref{eq:gamma} and \eqref{eq:eta} result from a Taylor expansion of the integrals $I_{++}(\omega_\qq,q)$, $I_{--}(\omega_\qq,q)$ 
and $I_{+-}(\omega_\qq,q)$ 
after replacement of $\hbar\omega_\qq$ with the expansion \eqref{eq:fleur}.
At each order, we reuse the results of the lower orders, that is the value \eqref{eq:vitesseson2} of $c$ to obtain $\gamma$, and then those of $c$ and 
$\gamma$ \eqref{eq:gamma} to obtain $\eta$.
We encounter integrals involving in the denominator high powers of $\epsilon_\kk$ (or of $[(u^2-x^{-1})^2+1]^{1/2}$ after the $k$-to-$u$ change of variable).
They can be evaluated by repeated integration by parts, as explained in Appendix \ref{app:integrales}.

\begin{figure}[htb]
\centerline{\includegraphics[width=0.45\textwidth,clip=]{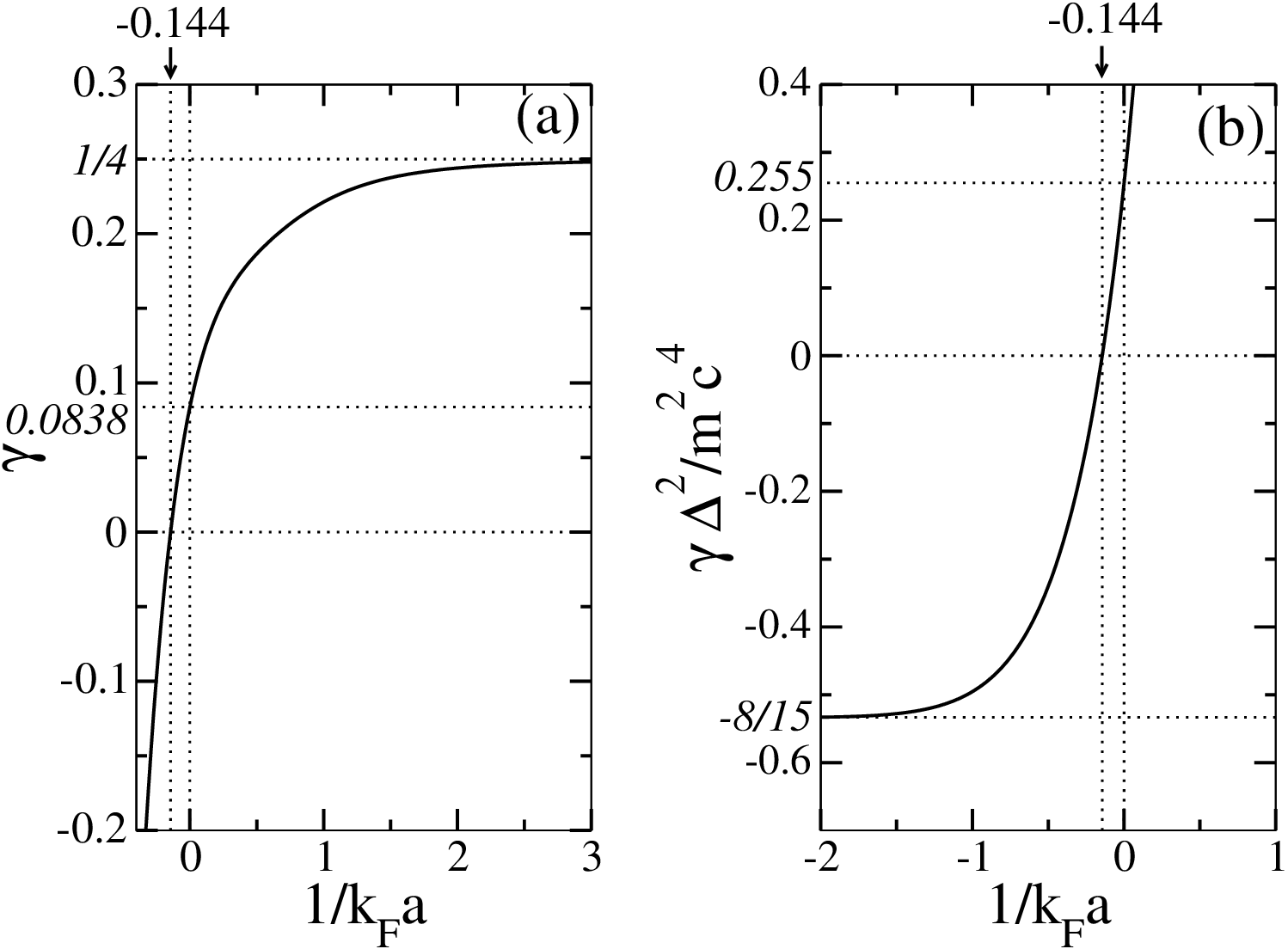}}
\centerline{\includegraphics[width=0.45\textwidth,clip=]{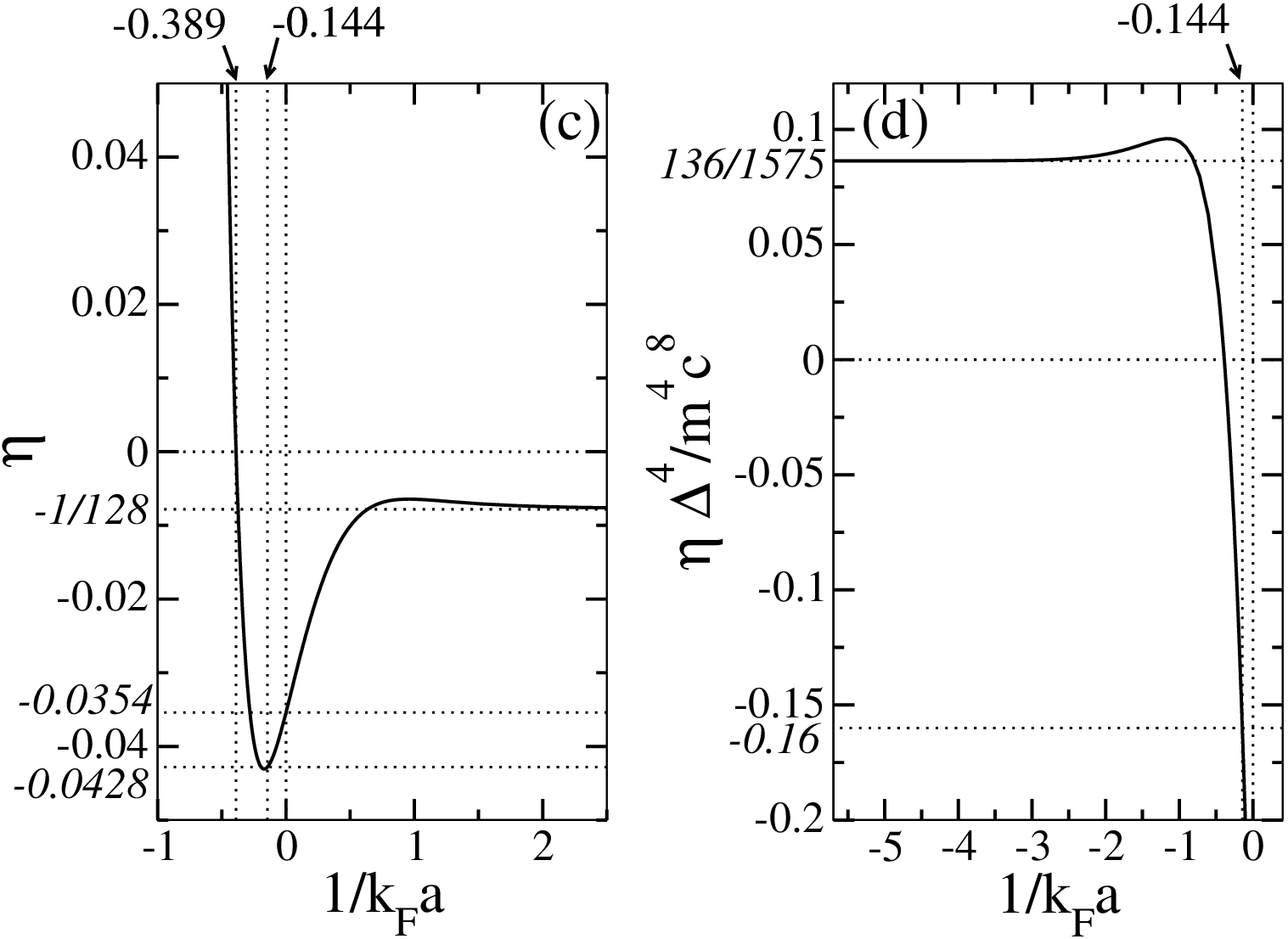}}
\caption{\label{fig:gammaeta} 
Dimensionless coefficients (a) $\gamma$ and (c) $\eta$ of the terms $q^3$ and $q^5$ in the RPA dispersion relation of the bosonic excitations
for $q\to 0$, see equation (\ref{eq:fleur}), as functions of the interaction parameter $1/(k_F a)$.
In the BEC limit, they have a finite limit corresponding to the Bogoliubov dispersion relation, see equations (\ref{eq:Bogo},\ref{eq:limCBE}).
In the BCS limit, they diverge, and it is more appropriate to express the wave number $q$ in units of $\Delta/(\hbar c)$, which amounts to considering
the quantities (b) $\gamma \Delta^2/(m^2c^4)$ and (d) $\eta \Delta^4/(m^4c^8)$ which have a finite limit.
In between these two limits, $\gamma$ and $\eta$ vanish and change sign for $1/(k_F a)\simeq-0.144$ and $1/(k_F a)\simeq-0.389$ respectively,
and they are weakly positive ($\gamma\simeq 0.0838$) or weakly negative ($\eta\simeq -0.0354$) at the unitary limit.
The value of $\eta$ is relevant mainly at the point where $\gamma$ vanishes. At this point, $\eta\simeq -0.0428$ so the RPA dispersion relation is concave
over a neighborhood of $q=0$.}
\end{figure}

We have plotted in Figure \ref{fig:gammaeta}  the coefficients $\gamma$ and $\eta$ as functions of the parameter $1/(k_F a)$ (which we have preferred here to $\Delta/\mu)$.
Let us briefly review their asymptotic behaviors in the BEC $1/(k_F a) \rightarrow +\infty$ and BCS $1/(k_F a) \rightarrow -\infty$ limits, and their
values in some specific relevant regimes.

%%%%%%%%%%%%%%%%%%% FIGURE %%%%%%%%%%%%%%%%%%%%%%%%%
%%%%%%%%%%%%%%%%%%% ENDFIGURE %%%%%%%%%%%%%%%%%%%%%%%%%
%%%%%%%%

\subsubsection{BEC limit}%%%%%%%% 
In the BEC limit $k_F a\rightarrow 0^+$, the system is equivalent to a weakly interacting gas of bosons of mass $2m$, with a chemical potential
\be
\mu_B=2\mu-E_{\rm dim},
\ee
where $E_{\rm dim}=-\hbar^2/(ma^2)$ is the internal energy of a dimer \cite{Leyronas2007,Leyronas2008}. 
The dispersion relation of the bosonic excitations is then known to be convex and to take the Bogoliubov form at chemical potential $\mu_B$
when $q=o(1/a)$~\cite{CKS2006}:
\be
\hbar\omega_\qq^{\rm Bog}=\left[\frac{\hbar^2q^2}{4m}\left(\frac{\hbar^2q^2}{4m}+2\mu_B\right)\right]^{1/2},
\label{eq:Bogo}
\ee
in which case the sound velocity is given by $2mc^2 = \mu_B$. The coefficients $\gamma$ and $\eta$ are thus expected to have the following limits:
\be
\gamma\underset{k_Fa\to 0^+}{\to}\frac{1}{4}\ \ \ \mbox{and}\ \ \ \eta\underset{k_Fa\to 0^+}{\to}-\frac{1}{128}.
\label{eq:limCBE}
\ee
This is confirmed by equations (\ref{eq:gamma}) and (\ref{eq:eta}) taken at the BEC limit, that is for
$x=O(k_F a)^{3/2}\to 0^-$ \cite{Varenna} and, as shown by equation (\ref{eq:dDeltadmu}) after the change of variable
$u=u'/|x|^{1/2}$,  for $y\sim -4/x$.

%%%%%%%%
\subsubsection{BCS limit}%%%%%%%% 
When $k_F a\rightarrow 0^-$, the lower border of the two-fermionic-excitation continuum (at fixed total wave number $q<q_{\rm sup}$)
becomes exponentially weak and forces the bosonic excitation branch, which cannot enter into this continuum, to bend downward \cite{CKS2006,Castin2015}.
This bending takes place over a wave number range $q_c$ such that $\hbar c q_c =\Delta$, that is such that the leading term in the expansion \eqref{eq:fleur}
is of the order of $\Delta$.  This means that the collective modes are affected by the pairs internal structure when the mode wavelength becomes comparable to the pairs size in real space $\approx \hbar^2 k_F/(m \Delta)$ \cite{Strinati1998}, a quantity that is indeed of order $1/q_c$ since $c\propto \hbar k_F/m$ in the BCS limit.
This qualitatively explains why the dispersion relation is a concave function of $q$ for low $q$ and low
$\Delta/\mu$ in Figure \ref{fig:conc}, and why $\gamma <0$ in the BCS limit. 

More quantitatively, we expect that the normalized energy  $\hbar \omega_\qq/\Delta$ becomes a universal function of $q/q_c=\hbar c q/\Delta$ when $k_F a\to 0^-$, in which case all the terms in square brackets in equation (\ref{eq:fleur}) are of the same order of magnitude for $q=q_c$, that is $|\gamma (\hbar q_c/mc)^2|\approx 1$, $|\eta(\hbar q_c/mc)^4|\approx 1$.
This is indeed what we find by taking the limits $x\to 0$ and $y\to 0$ in equations (\ref{eq:gamma}) and (\ref{eq:eta}):
\be
\gamma \underset{k_Fa\rightarrow 0^-}{\sim} -\frac{8}{15} \bb{\frac{mc^2}{\Delta}}^2 \ \ \mbox{and}\ \ 
\eta \underset{k_Fa\rightarrow 0^-}{\sim} \frac{136}{1575} \bb{\frac{mc^2}{\Delta}}^4,
\ee
the first result reproducing that of reference \cite{Strinati1998}.

%%%%%%%%
\subsubsection{Crossover region}%%%%%%%% 
In the crossover region between BEC and BCS, $\gamma$ is an increasing function of $1/(k_Fa)$. It vanishes and changes sign for the value $x_0$ of $x=\Delta/\mu$ given by
\be
x_0\simeq 0.868567.
\ee
This value corresponds to $1/(k_F a)\simeq-0.144292$, in agreement with the numerical result of Figure  \ref{fig:conc} and with reference
\cite{Salasnich2015}. 
The RPA prediction is then that the dispersion relation of a unitary gas is convex close to $q=0$:
\be
\gamma\underset{(k_F a)^{-1}=0}{\simeq} 0.083769.
\ee
The coefficient $\eta$ changes sign for a value $x_1$ of $\Delta/\mu$ given by
\be
x_1\simeq0.566411
\ee
corresponding to $1/(k_F a)\simeq-0.389027$. It is negative both at unitarity
\be
\eta\underset{(k_F a)^{-1}=0}{\simeq} -0.035416,
\ee
and at the point $x_0$ where $\gamma=0$:
\be
\eta(x_0) {\simeq} -0.042794.
\ee
At that very point the sign of $\eta$ is important as it determines the concavity of the dispersion relation close to $q=0$.
%%%%%%%%
\subsubsection{Locus of the inflection points}%%%%%%%% 
The coefficients $\gamma$ and $\eta$ allow us to find analytically the border between the orange and the red zones of 
Figure \ref{fig:conc} for small $q$, that is the ensemble of points with coordinates $(q_{\rm infl}/k_\mu,\Delta/\mu)$ such that the second derivative $\dd^2\omega_\qq/\dd q^2$ is zero. 
Using the expansion \eqref{eq:fleur} for $\omega_\qq$ and expanding the coefficients $\gamma(x)$ and $\eta(x)$  around $x=x_0$, to order one and order zero in $x-x_0$ respectively, one obtains the equation
\be
\frac{q_{\rm infl}^2}{k_\mu^2}
\underset{x\to x_0^+}{\sim}-\frac{3\gamma'(x_0)}{10\eta(x_0)}\frac{mc^2}{\mu}(x-x_0) \simeq 2.015858(x-x_0)
\label{eq:qinfl}
\ee
plotted as a black dashed curve in  Figure \ref{fig:conc}, which reaches the axis  $q=0$ with a horizontal tangent.

In contrast, the border between the red and orange zones in Figure \ref{fig:conc} reaches the border of the existence domain of the collective excitation branch with an
oblique tangent. This is due to the fact that the third derivative of $q\mapsto \omega_\qq$ is nonzero at the contact point $q=q_{\rm sup}$ contrarily to what happens at $q=0$.
 %%%%%%%%%%%%%%%%%%%%%%%%%%%%%%%%%%%%%%%%%%%%%

\section{Conclusion}

%%%%%%%%%%%%%%%%%%%%%%%%%%%%%%%%%%%%%%%%%%%%%
We have considered a spatially homogeneous unpolarized gas of spin-$1/2$ fermions at zero temperature, and
we have obtained analytically the spectrum $\hbar \omega_\qq$ of the bosonic excitation branch predicted by the RPA up to order $5$ included in the wave vector $q$ 
close to $q=0$. The coefficients of the obtained expansion are rational fractions of two variables
$\Delta/\mu$ and $(\partial\Delta/\partial\mu)_a$, where the second variable can be analytically related to the first one using the BCS equation of state. This allows us to show analytically that the dispersion relation  $q\mapsto \omega_\qq$ is concave close to $q=0$ when $1/(k_F a)$ is between $-\infty$ and a value close to $-0.144$, 
a point where the first correction to the linear dispersion relation is of order $q^5$ with a slightly negative coefficient.
For $-0.144<1/(k_F a)<0.157$ the branch is convex close to $q=0$ but becomes concave when $q$ increases, and it remains so for $q$ increasing up to the maximal possible value $q_{\rm sup}$ if $1/(k_F a)<0.022$, while it becomes convex again in the opposite case $0.022<1/(k_F a)<0.157$. Beyond  $1/(k_F a)=0.157$ the bosonic branch is convex over its whole existence domain.

A straightforward application of our quintic approximation (\ref{eq:fleur}) for the spectrum is to determine if
a low-$q$ collective excitation of the Fermi gas can decay {\sl via} a Beliaev process, that is into two collective excitations 
of wave vectors $\qq_1$ and $\qq_2=\qq-\qq_1$. Energy conservation allows such a process if $\omega_\qq > \Omega_\qq^{\rm inf}$
where $\Omega_\qq^{\rm inf}=\inf_{\qq_1} (\omega_{\qq_1}+\omega_{\qq-\qq_1})$ is the lower border of the two-excitation continuum
at fixed total wave vector $\qq$. If $\Delta/\mu$ is away from the critical value $x_0\simeq 0.869$ where the coefficient $\gamma$ of $q^3$ vanishes,
the dispersion relation is either entirely convex or entirely concave at low $q$, and the Beliaev decay is respectively allowed
or forbidden. If $\Delta/\mu$ is close to the critical value $x_0$, the dispersion relation has an inflection point at low $q$.
We then apply the analysis of reference \cite{Castin2015} to Eq.~(\ref{eq:fleur})
and we find $\Omega_\qq^{\rm inf}=\min(2\omega_{\qq/2},\omega_\qq)$
\footnote{Since the dispersion relation $q\mapsto\omega_\qq$ is here an increasing function, only the cases (i) and (ii)
of Eq.~(12) in reference \cite{Castin2015}, where $\qq_1$ and $\qq-\qq_1$ have the same direction 
at the minimum of $\omega_{\qq_1}+\omega_{\qq-\qq_1}$, need to be considered. 
It remains to minimize the function $f(q_1)=\omega(q_1)+\omega(q-q_1)$ over the interval $[0,q/2]$,
where $\omega(q)=\omega_\qq$ is given by Eq.~(\ref{eq:fleur}). 
Using the parametrization $q_1=(1-t)q/2$, with $t\in [0,1]$, we find that $f(q_1)$ is a concave parabolic function of $t^2$.
%Computing $f'(q_1)$ and $f''(q_1)$
%and using the parametrization $q_1=(1-t)q/2$, with $t\in [0,1]$, we find that $f'(q_1)$ has a root if and only if
%$5/12< \gamma/[|\eta|(\hbar q/mc)^2]< 5/6$, in which case $f''(q_1)$ is negative at the root. 
The minimum of $f(q_1)$
is thus reached at the border of the interval and is either $f(0)=\omega_\qq$ or $f(q/2)=2\omega_{\qq/2}$.}. 
The Beliaev decay is thus allowed if
\be
\left(\frac{\hbar q}{mc}\right)^2 < \frac{8\gamma}{5|\eta|}
\ee
to leading order in $\gamma$, that is in $\Delta/\mu-x_0$.
At low but nonzero temperature, there exist additional decay mechanisms: 
(i) The Landau mechanism $\qq+\qq_1\to \qq_2$ is forbidden whenever the Beliaev one is since it can be viewed
as an inverse Beliaev mechanism with an initial wave vector $\qq_2$ of modulus $>q$;
(ii) the higher-order Landau-Khalatnikov decay process $\qq+\qq_1\to \qq_2+\qq_3$ is always allowed, but it is subleading
when the Beliaev or Landau processes are present.

Our results on the concavity of the dispersion relation close to $q=0$ can be tested experimentally in a gas of cold atoms trapped in a flat bottom potential \cite{Hadzibabic2013}. This can be done either (i) indirectly by measuring dissipative effects such as the damping of collective excitations at low temperature, or dispersive effects 
such as the spreading of a wave packet of sound waves created by a laser pulse \cite{Ketterle1997,Grimm2013}, or (ii) directly by accessing the dispersion relation at low temperature {\sl via} the dynamic structure factor of the Fermi gas through Bragg excitation at a selected wave vector  $\qq$ \cite{Ketterle1999,Davidson2002,Hannaford2010}
\footnote{In reference \cite{Davidson2002} the spectrum is measured with an uncertainty $\pm 18$Hz at low $q$. For the typical Fermi 
temperature $T_F=1\mu$K, this corresponds to an uncertainty $\pm 2\times 10^{-3}$ on $\hbar\omega_\qq/\mu$ that is
on the vertical axis of Figure \ref{fig:disp}, indicating that the convexity of $q\mapsto \omega_\qq$ is experimentally 
determinable for the unitary Fermi gas.}.
Supplemented by kinetic equations for the collective mode occupation numbers, our results open the way to an analytical determination of the phase diffusion coefficient at low temperature, hence to the intrinsic limit to the coherence time of the condensate of pairs in a finite-size Fermi gas \cite{KCS2015}.

%%%%%%%%%%%%%%%%%%%%%%%%%%%%%%%%%%%%%%%%%%%%%

%%%%%%%%%%%%%%%%%%%%%%%%%%%%%%%%%%%%%%%%%%%%%
\appendix
%%%%%%%%%%%%%%%%%%%%%%%%%%%%%%%%%%%%%%%%%%%%%

\section{Expressing integrals in terms of the variables $x$ and $y$}
\label{app:integrales}
%%%%%%%%%%%%%%%%%%%%%%%%%%%%%%%%%%%%%%%%%%%%%
In the expansion of $I_{++}(\omega_\qq,q)$, $I_{--}(\omega_\qq,q)$ and $I_{+-}(\omega_\qq,q)$ at low $q$, and after a rescaling of the wave vectors by $k_\Delta$ as in equation \eqref{eq:dDeltadmu}, $k=u k_\Delta$, we encounter integrals of the form
\bea
I_{n,p} &=& \frac{k_\Delta^3}{2\pi^2\rho}\int_0^{+\infty}\dd u\frac{u^{2p+2}}{\epsilon_u^n},\label{eq:notationunifiee1} \\
J_{n,p} &=& \frac{k_\Delta^3}{2\pi^2\rho}\int_0^{+\infty}\dd u\frac{u^{2p+2}\xi_u}{\epsilon_u^n} \label{eq:notationunifiee2}
\eea
with $n\in2\mathbb{N}^*+1$, $p\in\mathbb{N}$, 
\bea
\xi_u     &=& u^2-\frac{1}{x}, \\
\epsilon_u&=& \sqrt{\xi_u^2+1},
\eea
and the total density $\rho$ is given by equation (\ref{eq:rho}).
The integrals giving $I_{n,p}$ and $J_{n,p}$ are convergent for $n-p\geq2$ and for $n-p\geq 3$, respectively.
Integrals that depend on the direction of $\kk$ can be expressed in the forms 
 \eqref{eq:notationunifiee1} and \eqref{eq:notationunifiee2} after angular integration:
\begin{multline}
\int \dd^3k f(k) \bb{\frac{\hbar^2\kk\cdot\qq}{m}}^{2p} = \\ 
\frac{4\pi}{1+2p} \bb{\frac{\hbar^2q^2}{m}}^{p} \int_0^{+\infty} \dd k k^2 f(k) \bb{\frac{\hbar^2 k^2}{m}}^{p} 
\end{multline}
where $f(k)$ is an arbitrary function of the modulus of $\kk$.

Let us first establish the four recurrence relations
\bea
I_{n,p} &=& \frac{n-3}{n-2}I_{n-2,p}-\frac{2p+1}{2(n-2)} J_{n-2,p-1}, \label{eq:Inp} \\
J_{n,p} &=& \frac{2p+1}{2(n-2)} I_{n-2,p-1}, \label{eq:Jnp} \\
I_{n,p} &=& J_{n,p-1}+\frac{I_{n,p-1}}{x}, \label{eq:Inpbis}\\
J_{n,p} &=&\frac{J_{n,p-1}}{x} +  I_{n-2,p-1}- I_{n,p-1}\label{eq:Jnpbis},
\eea
holding under the conditions $1\leq p\leq n-4$ for the first relation,  $1\leq p\leq n-3$ for the second one, $1\leq p\leq n-2$  for the third one 
and $1\leq p\leq n-3$ for the last one.
In order to derive the relation \eqref{eq:Inp}, we integrate by parts the integral 
\be
\frac{k_\Delta^3}{2\pi^2\rho}\int_0^{+\infty}\dd u\frac{u^{2p+2}\xi_u^2}{\epsilon_u^n}=I_{n-2,p}-I_{n,p},
\ee
selecting $u\mapsto u^{2p+1} \xi_u$ as the function to be differentiated. In order to derive the relation \eqref{eq:Jnp}, 
we integrate by parts the integral defining $J_{n,p}$ in equation  (\ref{eq:notationunifiee2}), selecting $u\mapsto u^{2p+1}$ as the function to be differentiated. 
In both cases, we note that the function $u\mapsto u \xi_u/\epsilon_u^n$ admits the primitive 
$u\mapsto -[2(n-2)\epsilon_u^{n-2}]^{-1}$.
Finally, we simply write $u^{2p+2}=u^{2p}(\xi_u+x^{-1})$  in the integrand of  (\ref{eq:notationunifiee1}) in order to obtain
(\ref{eq:Inpbis}),
and we write $u^{2p+2}\xi_u=u^{2p}(\epsilon_u^2-1+x^{-1}\xi_u)$ in the integrand of (\ref{eq:notationunifiee2}) in order to obtain (\ref{eq:Jnpbis}).
This procedure generalizes that of reference \cite{Strinati1998}.

We now show by induction using the relations (\ref{eq:Inp},\ref{eq:Jnp},\ref{eq:Inpbis},\ref{eq:Jnpbis})
that $I_{n,p}$ and $J_{n,p}$ can be expressed as functions of $I_{3,0}$ and $J_{3,0}$,
for all odd $n\geq 3$ and for all positive $p$ within the existence domain of the integrals.
Let $n$ be odd and $\geq 3$ and assume that we know all the $I_{n,p}$, $0\leq p\leq n-2$, and all the $J_{n,p}$, $0\leq p\leq n-3$.
Then (i) $I_{n+2,1}$ and $J_{n+2,1}$ can be deduced using (\ref{eq:Inp}) and (\ref{eq:Jnp}), (ii)
using (\ref{eq:Jnpbis}) and (\ref{eq:Inpbis}) we obtain a Cramer system for $I_{n+2,0}$ and $J_{n+2,0}$:
\bea
x^{-1} J_{n+2,0} - I_{n+2,0} &=& J_{n+2,1} - I_{n,0}, \\
J_{n+2,0} +x^{-1} I_{n+2,0} &=& I_{n+2,1},
\eea
which we solve, (iii) we use (\ref{eq:Inpbis}) and (\ref{eq:Jnpbis}) to access the values of $I_{n+2,p}$ and $J_{n+2,p}$ for $p\geq 2$. 
We set the induction basis at $n=3$, by expressing $I_{3,1}$ as a function of $I_{3,0}$ and $J_{3,0}$ due to (\ref{eq:Inpbis}).

Finally we relate $I_{3,0}$ and $J_{3,0}$ to $x$ and $y$ thanks to the relations \eqref{eq:dDeltadmu} and \eqref{eq:thetasurN} of the main text,
which take the form $y=J_{3,0}/I_{3,0}$ and $I_{3,0}=3x/[2(1+xy)]$ with the notation of this appendix.

%\bibliography{biblio}
%\bibliographystyle{unsrt}

\providecommand*\hyphen{-}

\end{document}

\title{Concavité de la branche d'excitation collective d'un gaz de fermions dans la zone de raccordement CBE-BCS}
\author{H. Kurkjian, Y. Castin, A. Sinatra}
\affiliation{Laboratoire Kastler Brossel, ENS-PSL, CNRS, UPMC-Sorbonne Universit\'es et Coll\`ege de France,
Paris, France}

\begin{abstract}
Nous étudions la concavité de la relation de dispersion $q\mapsto \omega_{\mathbf{q}}$
des excitations bosoniques d'un gaz tridimensionnel non polarisé de fermions de spin $1/2$ telle que décrite par l'Approximation de la Phase Aléatoire (RPA).
Dans la limite des faibles nombres d'onde $q$ nous obtenons le spectre analytiquement jusqu'à l'ordre $5$ en $q$. 
Au voisinage de $q=0$, un changement de concavité entre une branche concave dans la limite BCS et convexe dans la limite CBE se produit en 
$\Delta/\mu\simeq0,869$ ($1/(k_F a)\simeq-0,144$), où $a$ est la longueur de diffusion entre fermions de spins opposés, $k_F$ le nombre d'onde de Fermi
et $\Delta$ le {\sl gap} d'après la théorie BCS, et $\mu$ le potentiel chimique. 
En ce point, la branche est concave du fait d'un terme d'ordre $5$ négatif. 
Ces résultats sont complétés par une étude numérique qui montre l'évolution dans tout le plan $(q,\Delta)$
de la frontière entre la zone où $q\mapsto \omega_{\mathbf{q}}$ est concave et celle
où $q\mapsto \omega_{\mathbf{q}}$ est convexe.
\end{abstract}

% 03.75.Kk	Dynamic properties of condensates; collective and hydrodynamic excitations, superfluid flow
% 67.85.Lm	Degenerate Fermi gases
% 47.37.+q	Hydrodynamic aspects of superfluidity; quantum fluids
\maketitle

%%%%%%%%%%%%%%%%%%%%%%%%%%%%%%%%%%

\section{Introduction}

%%%%%%%%%%%%%%%%%%%%%%%%%%%%%%%%%%

Les gaz d'atomes froids permettent d'explorer le problème à $N$ corps avec une grande flexibilité dans le choix de ses paramètres microscopiques. 
En particulier, dans les gaz de fermions de spin $1/2$, on peut ajuster expérimentalement la force des interactions entre les particules grâce à des résonances de Feshbach sans induire de fortes pertes de particules \`a trois corps. 
Cette liberté, unique dans les systèmes fermioniques, a permis  aux expériences d'atomes froids \cite{Thomas2002,Salomon2003,Grimm2004,Grimm2004b,Ketterle2004,Ketterle2005,Salomon2010,Zwierlein2012,Grimm2013} 
d'étudier comment l'on passe 
continûment d'un superfluide de paires de Cooper dans le régime dit de Bardeen-Cooper-Schrieffer (BCS) à un superfluide de dimères quasi-bosoniques dans le régime de Condensation de Bose-Einstein (CBE). 
L'autre avantage des gaz d'atomes froids est de permettre une description théorique simple des interactions, qui ont lieu dans un régime dilué et de basse énergie. 
Pour un gaz de fermions dans deux états de spin $\uparrow$ et $\downarrow$, on montre ainsi que les seules interactions significatives à basse température et faible densité ont lieu dans l'onde $s$, entre fermions de spins opposés, et se caractérisent par un unique paramètre appelé longueur de diffusion et noté $a$. 
Ce cadre théorique avantageux autorise des études entièrement analytiques de propriétés expérimentalement accessibles du gaz en interaction, dans lesquelles cet article 
s'inscrit.

\`A température nulle, notre gaz tridimensionnel de fermions spatialement homogène non polarisé est enti\`erement appari\'e, et son
spectre d'excitation se compose de deux branches~: une branche fermionique d'excitation 
des degrés de liberté internes des paires de fermions $\uparrow\downarrow$ et une branche bosonique d'excitation 
de leur centre de masse, branche qui devient phononique dans la limite des grandes longueurs d'onde. 
Cette dernière branche est aussi dite collective car elle met en jeu un grand nombre de modes fermioniques d'excitation interne des paires. 
La branche fermionique est décrite à l'ordre le plus bas par la théorie BCS~;
la branche bosonique, elle, est abordée par différentes approches~: 
l'Approximation de la Phase Aléatoire (Random Phase Approximation, RPA) d'Anderson \cite{Anderson1958}, 
une approximation gaussienne de l'action dans un cadre d'intégrale de chemin \cite{Strinati1998,Randeria2014}, 
une approche par les fonctions de Green associée à une approximation diagrammatique \cite{CKS2006}, et une lin\'earisation des \'equations variationnelles BCS
d\'ependant du temps \cite{KCS2015f}.
Ces différentes théories, toutes approchées, convergent pour donner le même spectre d'excitation bosonique, décrit par la même équation implicite.

La concavité de ce spectre a été étudiée dans la limite BCS \cite{Strinati1998} et de manière qualitative dans la zone de raccordement CBE-BCS \cite{CKS2006,Castin2015,Salasnich2015}. 
Il manque donc une étude quantitative exhaustive, et c'est ce qui fait l'objet de cet article. 
En particulier, nous obtenons analytiquement le spectre jusqu'à l'ordre $5$ en le nombre d'onde $q$ du centre de masse des paires et nous en d\'eduisons
la concavité de la branche d'excitation sur un voisinage de $q=0$ dans toute la zone de raccordement CBE-BCS.

L'intérêt physique de notre étude est de déterminer quels processus d'amortissement des modes collectifs dominent à basse température. 
Si la branche est convexe sur un voisinage de $q=0$ alors ce sont les processus d'interaction entre phonons $2\,\mbox{phonons} \leftrightarrow 1\,\mbox{phonon}$ de Beliaev-Landau \cite{Beliaev1958,Shlyapnikov1998} qui dominent, 
tandis que si elle est concave ce sont les processus $2\,\mbox{phonons}\leftrightarrow2\,\mbox{phonons}$ de Landau-Khalatnikov \cite{Khalatnikov1949}.
En effet, on montre à partir de la conservation de l'énergie et de la quantité de mouvement des trois excitations mises en jeu que les processus Beliaev-Landau 
sont interdits pour une branche d'excitation concave. 
\`A basse température, la branche fermionique, dont le spectre possède une bande interdite, ne participe pas à l'amortissement des modes collectifs, 
sa contribution étant exponentiellement réduite \cite{Liu2011}. 
En outre, la connaissance quantitative du paramètre de concavité $\gamma$ permet de prédire le taux d'amortissement des phonons par les processus $2\,\mbox{phonons}\leftrightarrow1\,\mbox{phonon}$ 
dans le cas convexe au-del\`a de l'hydrodynamique quantique \cite{Salasnich2015}, ou $2\,\mbox{phonons}\leftrightarrow2\,\mbox{phonons}$ dans le cas concave où l'interaction effective pr\'edite par l'hydrodynamique quantique
fait intervenir des processus $2\,\mbox{phonons}\leftrightarrow1\,\mbox{phonon}$ virtuels et d\'epend de $\gamma$ \cite{Khalatnikov1949}. 
La connaissance de $\gamma$ donne enfin accès au coefficient de diffusion de la phase du condensat de paires, processus fondamental qui limite le temps de cohérence intrinsèque du gaz \cite{KCS2015f}.

%%%%%%%%%%%%%%%%%%%%%%%%%%%%%%%%%%

\section{\'Equation de la RPA sur le spectre d'excitation}

%%%%%%%%%%%%%%%%%%%%%%%%%%%%%%%%%%

L'équation de la RPA donnant implicitement l'énergie $\hbar \omega_\qq$ des excitations collectives en fonction du vecteur d'onde $\qq$ s'écrit~:
\be
I_{++}(\omega_\qq,q) I_{--}(\omega_\qq,q) = {\hbar^2\omega^2_\qq} \left[I_{+-}(\omega_\qq,q)\right]^2.
\label{eq:dedispersion}
\ee
Les quantités $I_{\sigma\sigma'}$ traduisent le caractère collectif des modes bosoniques~; 
ce sont des intégrales sur le vecteur d'onde relatif $\kk$ des paires, fonctions de $\omega_\qq^2$ {\sl via} le dénominateur de leur intégrande \cite{CKS2006}~:
\begin{widetext}
\bea
I_{++}(\omega,q)&=&\int \dd^3k\left[\frac{(\epsilon_{\kk+\qq/2}+\epsilon_{\kk-\qq/2})(U_{\kk+\qq/2}U_{\kk-\qq/2}+V_{\kk+\qq/2}V_{\kk-\qq/2})^2}{\hbar^2\omega^2
-(\epsilon_{\kk+\qq/2}+\epsilon_{\kk-\qq/2})^2}+\frac{1}{2\epsilon_\kk}\right] \\
I_{--}(\omega,q) &=&\int \dd^3k\left[\frac{(\epsilon_{\kk+\qq/2}+\epsilon_{\kk-\qq/2})(U_{\kk+\qq/2}U_{\kk-\qq/2}-V_{\kk+\qq/2}V_{\kk-\qq/2})^2}
{\hbar^2\omega^2-(\epsilon_{\kk+\qq/2}+\epsilon_{\kk-\qq/2})^2}+\frac{1}{2\epsilon_\kk}\right] \\
I_{+-}(\omega,q)&=&\int \dd^3k\frac{(U_{\kk+\qq/2}U_{\kk-\qq/2}+V_{\kk+\qq/2}V_{\kk-\qq/2})(U_{\kk+\qq/2}U_{\kk-\qq/2}-V_{\kk+\qq/2}V_{\kk-\qq/2})}{\hbar^2\omega^2-(\epsilon_{\kk+\qq/2}+\epsilon_{\kk-\qq/2})^2}.
\eea
\end{widetext}
Nous avons introduit ici les amplitudes $U_\kk$ et $V_\kk$ et les énergies propres $\epsilon_\kk$ des modes propres de la théorie BCS \cite{BCS1957}~:
\bea
\epsilon_\kk &=& \sqrt{\bb{\frac{\hbar^2k^2}{2m}-\mu}^2+\Delta^2} \\
U_{\kk} &=& \sqrt{\frac{1}{2}\bb{1+\frac{\frac{\hbar^2k^2}{2m}-\mu}{\epsilon_\kk}}} \\
V_{\kk} &=& \sqrt{\frac{1}{2}\bb{1-\frac{\frac{\hbar^2k^2}{2m}-\mu}{\epsilon_\kk}}}
\eea
où $m$ est la masse d'un fermion. Les deux paramètres naturels de la théorie BCS, qui nous serviront aussi à exprimer 
l'énergie $\hbar\omega_\qq$, sont le potentiel chimique $\mu$ du gaz, commun aux deux états de spin, et le \textit{gap} $\Delta$, 
largeur de la bande interdite dans le spectre BCS des excitations fermioniques lorsque $\mu$ est positif. 
Ils peuvent à loisir être échangés avec la longueur de diffusion $a$ et la densité totale du gaz $\rho$ en inversant les deux relations \cite{Varenna,KCS2013}~:
\bea
 \frac{m}{4 \pi \hbar^2 a}  &=& \int \frac{d^3k}{(2\pi)^3} \bb{ \frac{m}{\hbar^2 k^2}  -\frac{1}{2\epsilon_\kk}} \label{eq:g0} \\
 \rho                                     &=&  \int \frac{d^3k}{(2\pi)^3} 2|V_\kk|^2. \label{eq:rho}
\eea
En pratique, plut\^ot que la densit\'e, on utilise souvent le nombre d'onde de Fermi $k_F$ d\'efini par
$\rho = k_F^3/(6\pi^2)$.

%%%%%%%%%%%%%%%%%%%%%%%%%%%%%%%%%%

\section{\'Etude numérique globale de la concavité}

\begin{figure}[htb]
\includegraphics[width=0.4\textwidth,clip=]{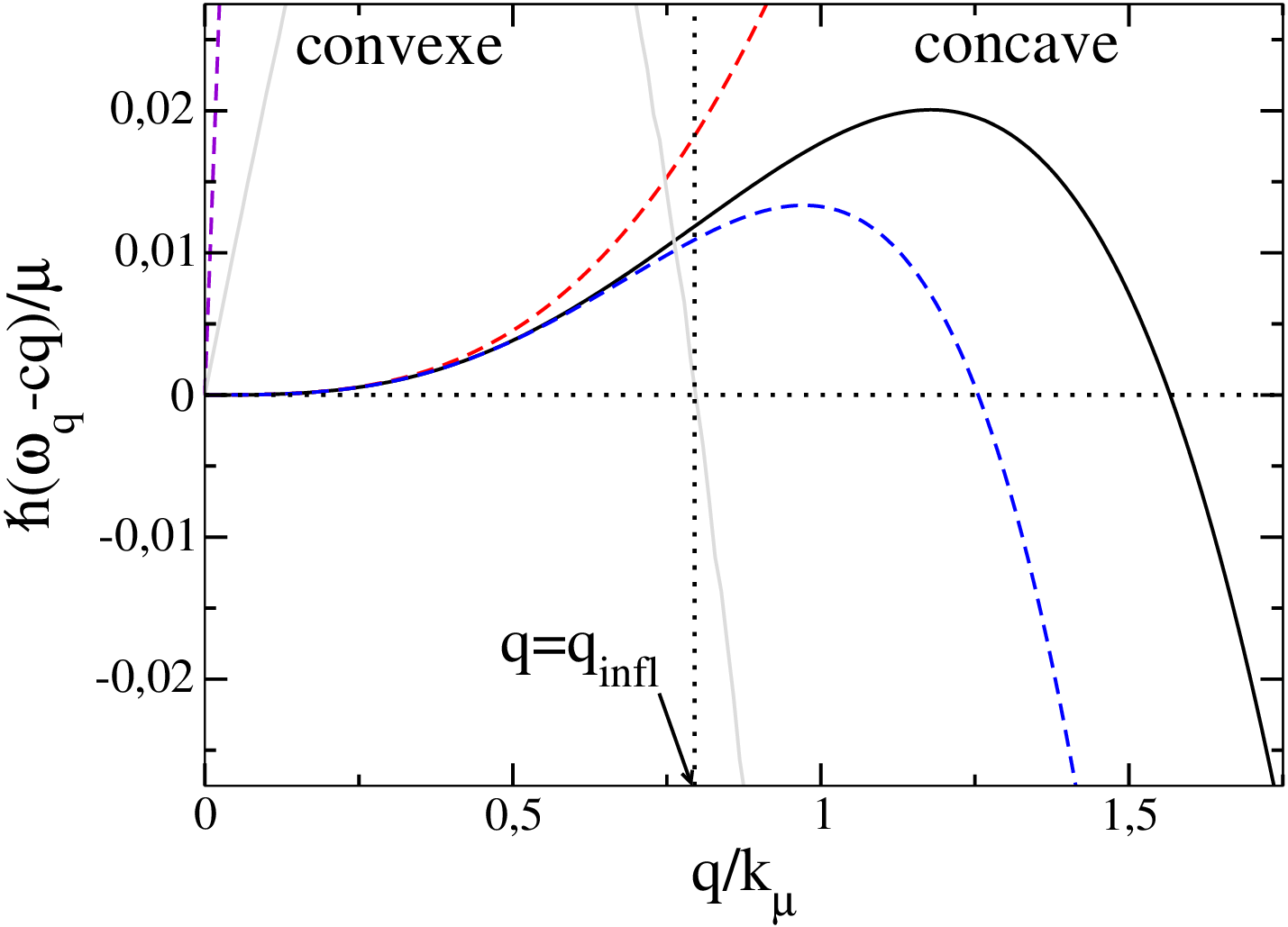}
\caption{
À l'unitarité $(k_F a)^{-1}=0$, c'est-à-dire pour $\Delta_0/\mu=1,162$, nous représentons en trait noir plein la relation de dispersion des excitations collectives {\sl après soustraction} de sa partie phononique $cq$ afin de mettre en \'evidence ses propriétés de concavité. Le nombre d'onde $q$ est en unités de $k_\mu=(2m\mu)^{1/2}/\hbar$.
La partie linéaire $q\mapsto cq$ (tireté violet) est représentée à titre de comparaison, de même que l'approximation cubique (tireté rouge sup\'erieur) et quintique (tireté bleu inf\'erieur) de $\omega_\qq-cq$ (voir plus loin l'équation (\ref{eq:fleur})). La dérivée seconde $\dd^2\omega_{\qq}/\dd q^2$, représentée en trait gris plein, s'annule au point d'inflexion $q_{\rm infl}\simeq 0,795 k_\mu$ marqué d'une ligne pointillée verticale. À gauche (respectivement à droite) de ce point, la relation de dispersion est convexe (respectivement concave).
\label{fig:disp}}
\end{figure}

En résolvant numériquement l'équation de dispersion \eqref{eq:dedispersion} nous obtenons la relation de dispersion 
$q\mapsto \omega_\qq$ sur tout son domaine d'existence. Nous donnons sur la figure \ref{fig:disp} l'exemple de la limite unitaire $1/(k_F a)=0$,
où $\Delta_0/\mu=1,162$ d'après la théorie BCS. Plutôt que la fonction $q\mapsto \omega_\qq$ elle-même, nous traçons en trait noir plein
la fonction $q\mapsto\omega_\qq-cq$, où $c$ est la vitesse du son et $q\mapsto cq$ la partie linéaire du spectre.
Les propriétés de concavité de cette fonction sont les mêmes que celles de $q\mapsto\omega_\qq$ mais se voient mieux graphiquement aux faibles $q$ car elles ne sont pas masquées par la partie linéaire. Cette dernière, bien que dominante, ne joue aucun rôle dans la détermination des processus d'amortissement décrits dans l'introduction. Sur la figure, on distingue nettement le point d'inflexion de la dispersion de relation à $q/k_\mu\simeq 0,795$, où $k_\mu=(2m\mu)^{1/2}/\hbar$ ; il sépare un intervalle aux faibles $q$ où la relation de dispersion est convexe d'un intervalle aux grands $q$ où elle est concave.
Par conséquent, d'après la RPA, les processus d'amortissement des excitations collectives du gaz dominants à basse température sont les processus $2\,\mbox{phonons} \leftrightarrow 1\,\mbox{phonon}$ de Beliaev-Landau.

\begin{figure}[htb]
\includegraphics[width=0.49\textwidth,clip=]{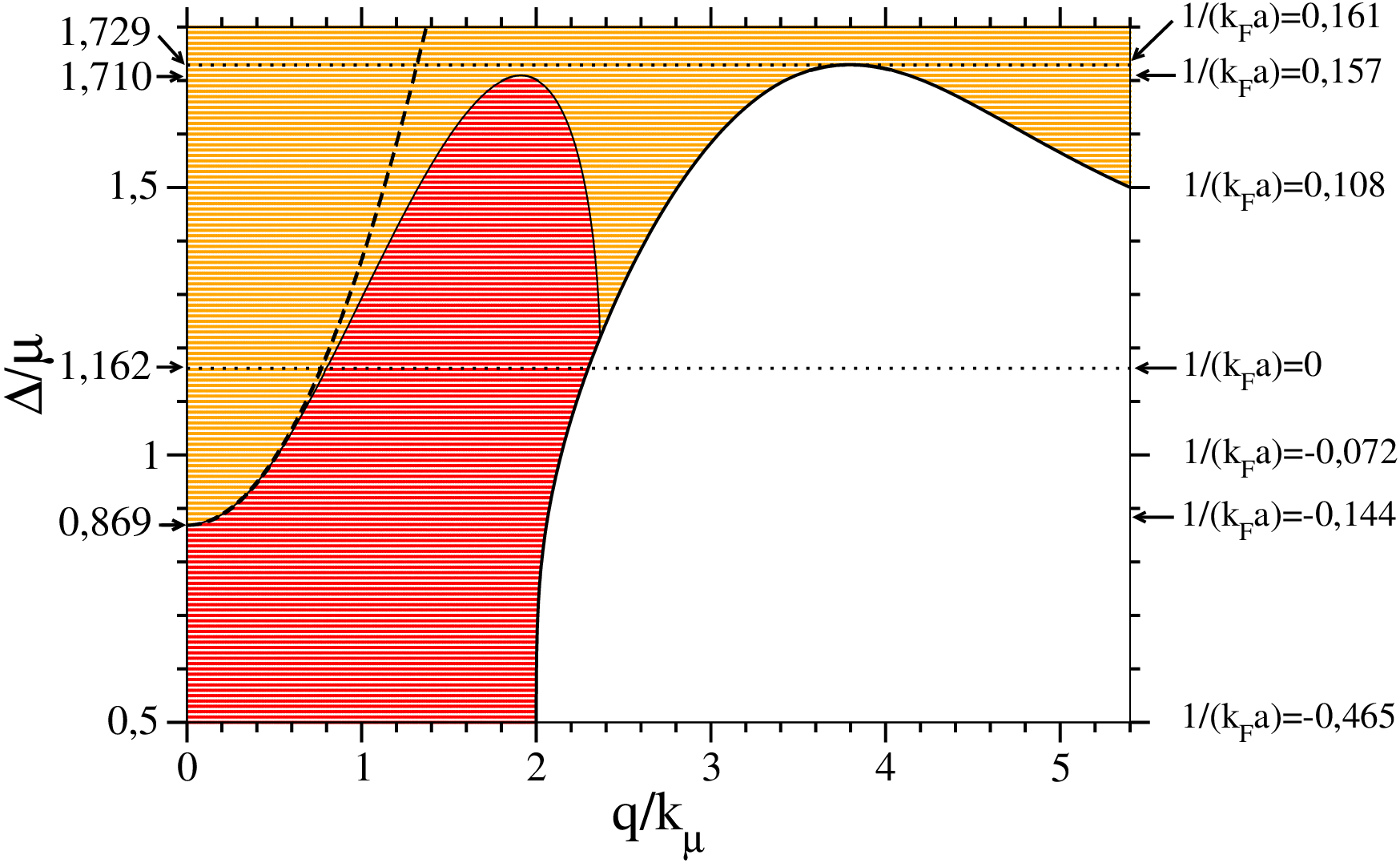}
\caption{Concavité locale de la branche bosonique $q\mapsto \omega_\qq$
selon les valeurs de $\Delta/\mu$ et du nombre d'onde $q$ mesuré en unités de $k_\mu=(2m\mu)^{1/2}/\hbar$. Les valeurs de $1/(k_F a)$ correspondant 
à celles de $\Delta/\mu$ sont indiquées sur le bord droit du graphe. Les points rouges correspondent à une branche localement concave, les points orange à une branche localement convexe. 
La ligne \'epaisse en trait plein délimite le domaine d'existence des solutions de l'équation de dispersion \eqref{eq:dedispersion}. La ligne noire tiretée est une prédiction analytique à faible $q$ 
(\textit{cf.} l'\'equation (\ref{eq:qinfl}) de la section \ref{sec:analytique}) 
de la frontière entre les zones rouge et orange, lieu des points d'inflexion de la courbe $q\mapsto\omega_\qq$ repr\'esent\'e par un trait fin. 
L'ordonnée $\Delta/\mu=0,869$ du point  où cette ligne rejoint l'axe $q=0$ et l'ordonnée $\Delta/\mu=1,710$ au-delà de laquelle la zone de concavit\'e disparaît sont indiquées par des flèches.
Enfin les deux lignes pointillées, dont les ordonnées sont elles aussi indiquées par des flèches, encadrent l'intervalle des valeurs de $\Delta/\mu$ ou $1/(k_F a)$ sur lequel le domaine d'existence en $q$ de la 
solution de l'équation \eqref{eq:dedispersion} n'est pas simplement connexe.
La ligne pointillée inférieure correspond à la limite unitaire, où la longueur de diffusion diverge $|a|\to+\infty$. \label{fig:conc}}
\end{figure}

%%%%%%%%%%%%%%%%%%%%%%%%%%%%%%%%%%
La figure \ref{fig:conc} synthétise les résultats sur la concavité de la branche bosonique que nous avons obtenus par résolution numérique de l'équation de dispersion \eqref{eq:dedispersion}. 
Plusieurs domaines se dégagent en fonction de la valeur de $\Delta/\mu$, ou s'il on préfère, de $1/(k_F a)$. $(i)$ Lorsque la longueur de diffusion est négative $a<0$, 
autrement dit pour $\Delta/\mu<1,162$ (pointillé inférieur), le domaine d'existence des solutions de l'équation \eqref{eq:dedispersion} en fonction de $q$ est compact et simplement connexe \cite{CKS2006}, de la forme $[0,q_{\rm sup}]$. 
La relation de dispersion est entièrement concave pour $\Delta/\mu<0,869$, puis, lorsque $0,869<\Delta/\mu<1,162$, elle est d'abord convexe aux faibles $q$ puis concave. 
Entre ces deux zones elle présente donc un point d'inflexion dont nous calculons analytiquement la position $q_{\rm infl}(\Delta/\mu)$ aux faibles $q$ (voir le tireté et 
l'\'equation (\ref{eq:qinfl}) de la section \ref{sec:analytique}). $(ii)$ 
De l'autre côté de la résonance ($a>0$), et jusqu'à $\Delta/\mu=1,729$ (pointillé supérieur), le domaine d'existence des solutions de l'équation \eqref{eq:dedispersion} se sépare en deux composantes connexes $[0,q_{\rm sup}]$ et $[q_{\rm inf},+\infty[$ \cite{CKS2006}. 
Tandis que la branche est toujours convexe dans la seconde composante $[q_{\rm inf},+\infty[$, d'intéressantes variations apparaissent dans la première composante $[0,q_{\rm sup}]$~: 
dans le sens des $q$ croissants, la branche est convexe 
puis concave pour $1,162<\Delta/\mu<1,22$, convexe puis concave puis à nouveau convexe pour $1,22<\Delta/\mu<1,710$ et enfin entièrement convexe pour $1,710<\Delta/\mu<1,729$. 
$(iii)$ Lorsque $\Delta/\mu$ est supérieur à $1,729$, ou est négatif,
les deux composantes du domaine d'existence se rejoignent, si bien qu'une solution à l'équation \eqref{eq:dedispersion} existe pour tout $q$ \cite{CKS2006}. 
La branche est alors entièrement convexe. 

Toutes ces valeurs numériques sont prédites par la RPA ou la th\'eorie BCS, donc sont approchées. 
La valeur de $\Delta/\mu$ \`a la limite unitaire est la seule que l'on puisse
\`a ce jour comparer aux exp\'eriences~: les r\'esultats exp\'erimentaux $\Delta\simeq 0,44 \hbar^2 k_F^2/(2m)$ \cite{Ketterle2008} et $\mu\simeq 0,376 \hbar^2 k_F^2/(2m)$
\cite{Zwierlein2012} conduisent \`a $\Delta/\mu\simeq 1,17$, remarquablement proche de la pr\'ediction de la th\'eorie BCS $\Delta/\mu\simeq 1,162$.
Rappelons aussi que le spectre de la RPA est issu d'un traitement lin\'earis\'e des fluctuations quantiques du champ de paires, qui n\'eglige les interactions
entre les quasi-particules bosoniques. Dans la r\'ealit\'e, ces interactions vont donner lieu \`a un d\'eplacement des \'energies propres $\hbar\omega_\qq$~;
elles vont aussi faire appara\^\i tre, m\^eme \`a temp\'erature nulle, une partie imaginaire dans $\omega_\qq$, d\'ecrivant la dur\'ee de vie finie des excitations, lorsque 
la concavit\'e de la relation de dispersion permet d'avoir des processus de Beliaev $1\,\mbox{phonon}\to 2\, \mbox{phonons}$ r\'esonnants \cite{Salasnich2015}.

%%%%%%%%%%%%%%%%%%%%%%%%%%%%%%%%%%

\section{\'Etude analytique de la concavité dans la limite des grandes longueurs d'onde}
\label{sec:analytique}
%%%%%%%%%%%%%%%%%%%%%%%%%%%%%%%%%%
La relation de dispersion peut être obtenue analytiquement dans la limite des grandes longueurs d'onde $q\rightarrow0$. Pour cela nous d\'eveloppons l'énergie 
propre du mode collectif jusqu'à l'ordre $5$ en $q$~:
\be
\hbar \omega_\qq \underset{q\to 0}{=} \hbar c q \left[1+\frac{\gamma}{8} \bb{\frac{\hbar q}{mc}}^2+\frac{\eta}{16} \bb{\frac{\hbar q}{mc}}^4 + O\bb{\frac{\hbar q}{mc}}^6
\right]. \label{eq:fleur}
\ee
\`A l'ordre le plus bas, comme pour tout syst\`eme superfluide, l'énergie est phononique avec une vitesse du son $c$ donnée par la formule hydrodynamique
\be
mc^2=\rho \bb{\frac{\partial\mu}{\partial\rho}}_a,\label{eq:vitesseson}
\ee
où la dérivée est prise à longueur de diffusion $a$ fixée comme le suggère la notation. Appliquée à l'équation d'état approch\'ee \eqref{eq:rho}, 
l'expression hydrodynamique \eqref{eq:vitesseson} 
permet d'obtenir la vitesse du son pr\'edite par la RPA, comme l'a \'etabli la r\'ef\'erence \cite{CKS2006} par d\'eveloppement de 
la solution $\omega_\qq$ de l'\'equation (\ref{eq:fleur}) au premier ordre en $q$. Nous en proposons ici une expression explicite sous forme d'une fraction rationnelle
\be
\frac{mc^2}{\mu} = \frac{2(xy+1)}{3(y^2+1)},\label{eq:vitesseson2}
\ee
en les variables 
\be
x = \frac{\Delta}{\mu} \ \ \ \ \mbox{et}\ \ \ \ 
y = \bb{\frac{\partial\Delta}{\partial\mu}}_a.
\ee
La variable $y$ s'exprime à son tour en fonction de $x$ grâce à la relation
\be
y =\frac{\int \dd^3 k \frac{\hbar^2k^2/(2m)-\mu}{\epsilon_\kk^3}}{\int \dd^3 k\frac{\Delta}{\epsilon_\kk^3}}   =\frac{\int_0^{+\infty} \dd u\frac{u^2(u^2-x^{-1})}{[(u^2-x^{-1})^2+1]^{3/2}}}{\int_0^{+\infty} \dd u \frac{u^2}{[(u^2-x^{-1})^2+1]^{3/2}}} 
\label{eq:dDeltadmu}
\ee
qui s'obtient en dérivant l'équation \eqref{eq:g0} par rapport à $\mu$ à longueur de diffusion $a$ fixée et en exprimant les vecteurs d'onde en unit\'es
de $k_\Delta=(2m\Delta)^{1/2}/\hbar$ pour faire apparaître 
la variable d'intégration $u$ sans dimension. 
Les intégrales apparaissant au numérateur et au dénominateur du troisième membre de \eqref{eq:dDeltadmu}
peuvent s'exprimer en termes d'intégrales elliptiques complètes de première et deuxième espèce \cite{Strinati1998}. 
Notons qu'\`a la limite unitaire $y=x$, puisque $\Delta$ et $\mu$ sont proportionnels par invariance d'\'echelle.

Pour obtenir l'expression \eqref{eq:vitesseson2} de la vitesse du son adimensionnée, nous dérivons l'équation d'état \eqref{eq:rho} par rapport à $\mu$ 
\`a $a$ fix\'e et exprimons toutes 
les intégrales utiles en fonction de $x$ et $y$ 
à l'aide de \eqref{eq:dDeltadmu} et de la relation 
\be
\frac{\int \frac{\dd^3 k}{(2\pi)^3} \frac{\Delta^3}{\epsilon_\kk^3}}{\rho}=\frac{\int_0^{+\infty} \dd u\frac{u^2}{[(u^2-x^{-1})^2+1]^{3/2}}}{\int_0^{+\infty} \dd u\, u^2\bb{1-\frac{u^2-x^{-1}}{[(u^2-x^{-1})^2+1]^{1/2}}}}
=\frac{3x/2}{1+xy},
\label{eq:thetasurN}
\ee
qui se démontre en intégrant par parties le dénominateur du second membre, $u^2$ étant la fonction à intégrer.

Pour obtenir les coefficients sans dimension $\gamma$ et $\eta$ des ordres supérieurs en $q$ de $\hbar\omega_\qq$ dans l'\'equation \eqref{eq:fleur},
on ne peut s'appuyer sur des expressions thermodynamiques connues et on n'\'echappe donc pas au d\'eveloppement laborieux
de l'\'equation (\ref{eq:dedispersion}) en puissances de $q$.  Le r\'esultat peut s'écrire tout de même sous la forme de fractions 
rationnelles en $x$ et $y$~: 
\bea
\gamma           &=& \frac{\sum_{i=0}^4 P_i(y) x^i}{135 x^2 \left(x^2+1\right) \left(y^2+1\right)^3} \label{eq:gamma} \\
\eta             &=& \frac{\sum_{i=0}^8 Q_i(y) x^i}{1020600 \left(y^2+1\right)^6 x^4\left(x^2+1\right)^2}. \label{eq:eta}
\eea
Les polynômes $P_i(y)$ et $Q_i(y)$ qui apparaissent aux numérateurs ordonnés par puissances croissantes de $x$  des expressions de $\gamma$ et de $\eta$ valent~:
\bea
P_0(y) &=& -4 \left(13 y^4+16 y^2+8\right) \notag \\
P_1(y) &=& 4 y \left(13 y^4+41 y^2+8\right) \notag \\
P_2(y) &=& 50 y^6-21 y^4-252 y^2-61\\
P_3(y) &=& 2 y \left(y^4+32 y^2+71\right) \notag \\
P_4(y) &=& 35 y^6+56 y^4-13 y^2-54 \notag 
\eea
\begin{widetext}
\bea
Q_0(y) &=& 16 \left(7745 y^8+19528 y^6+20304 y^4+8384 y^2+1088\right) \notag \\
Q_1(y) &=& 32 y \left(2857 y^8+67 y^6-3186 y^4-7920 y^2-2624\right) \notag \\
Q_2(y) &=& -8 \left(12882 y^{10}+28061 y^8-26936 y^6+7221 y^4-24496 y^2-5232\right) \notag \\
Q_3(y) &=& -8 y \left(8456 y^{10}-9859 y^8+9977 y^6+145295 y^4+3523 y^2+23720\right) \notag \\
Q_4(y) &=& -17500 y^{12}-247996 y^{10}-1249743 y^8-1341332 y^6+337202 y^4-694392 y^2+18321 \\
Q_5(y) &=& -4 y \left(25564 y^{10}+36027 y^8-66984 y^6+92206 y^4+387932 y^2-56121\right) \notag \\
Q_6(y) &=& -2 \left(12250 y^{12}+115637 y^{10}+558246 y^8+1071518 y^6+589478 y^4-248499 y^2+53082\right) \notag \\
Q_7(y) &=& -4 y \left(12957 y^{10}+33764 y^8-41904 y^6-173106 y^4-96189 y^2+53406\right) \notag \\
Q_8(y) &=& -8575 y^{12}-44544 y^{10}-149742 y^8-360644 y^6-477615 y^4-270756 y^2-20412. \notag 
\eea
\end{widetext}
Nous avons obtenu les expressions analytiques \eqref{eq:gamma} et \eqref{eq:eta} par développement limité des intégrales $I_{++}(\omega_\qq,q)$, $I_{--}(\omega_\qq,q)$ 
et $I_{+-}(\omega_\qq,q)$ 
après remplacement de $\hbar\omega_\qq$ par le développement \eqref{eq:fleur}.
Nous réutilisons à chaque ordre les résultats des ordres inférieurs~: la valeur \eqref{eq:vitesseson2} de $c$ pour obtenir $\gamma$, puis celles de $c$ et de $\gamma$ \eqref{eq:gamma} pour obtenir $\eta$. 
Au cours du calcul, nous rencontrons des intégrales avec 
des puissances élevées de $\epsilon_\kk$ (ou $[(u^2-x^{-1})^2+1]^{1/2}$ après adimensionnement) au dénominateur~; 
leur calcul par intégrations par partie répétées est expliqué dans l'annexe \ref{app:integrales}. 

\begin{figure}[htb]
\centerline{\includegraphics[width=0.45\textwidth,clip=]{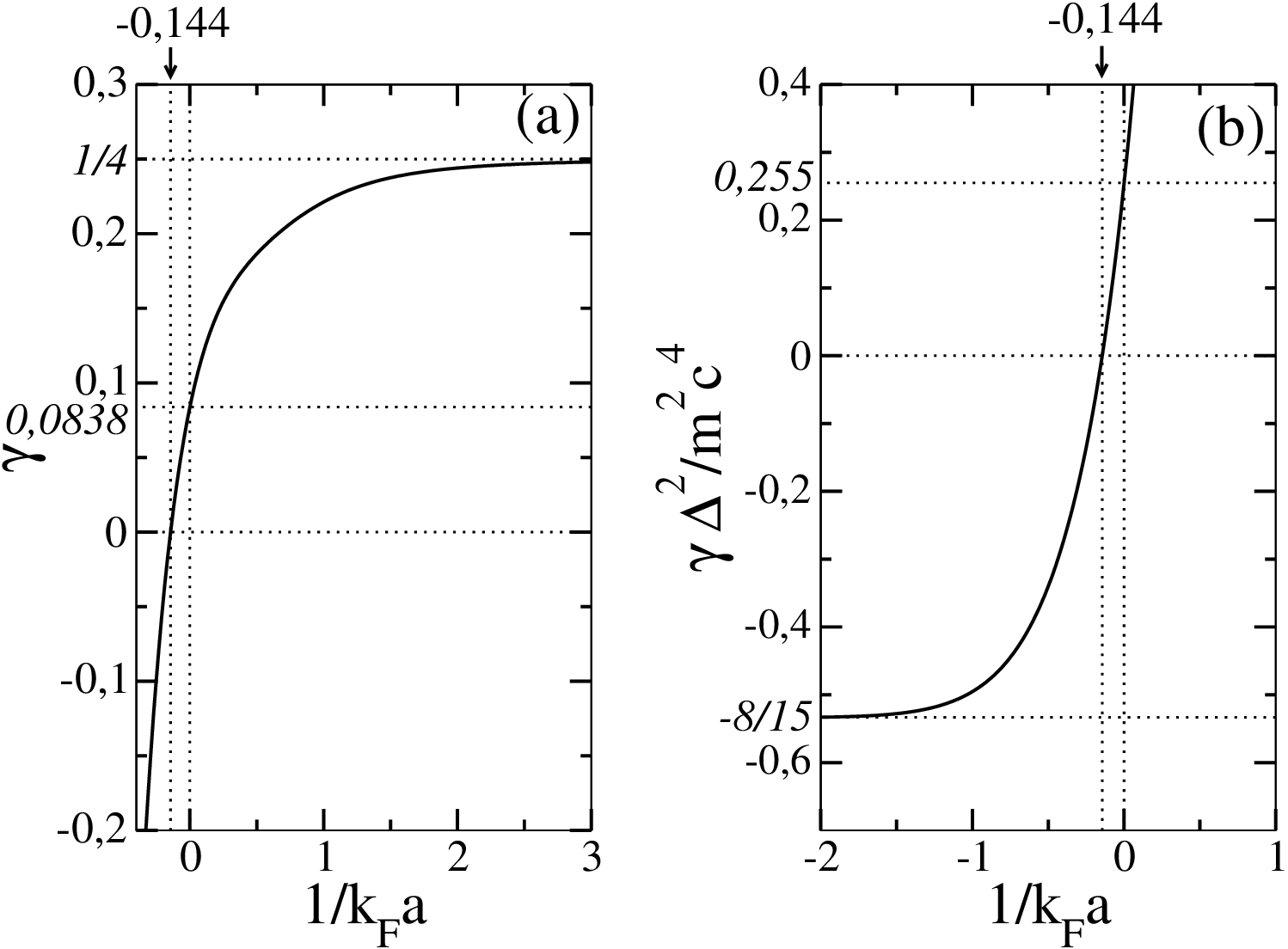}}
\centerline{\includegraphics[width=0.45\textwidth,clip=]{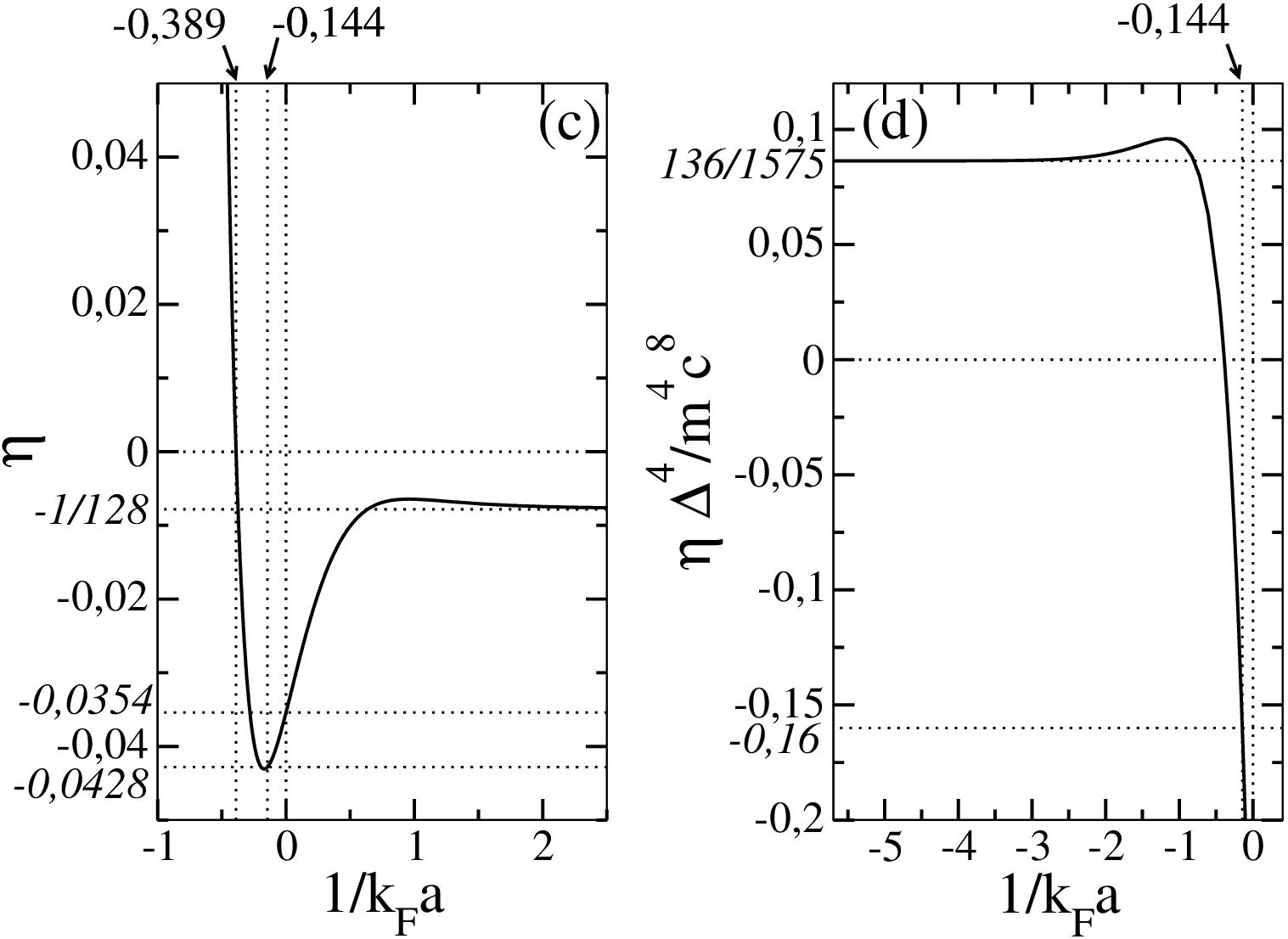}}
\caption{\label{fig:gammaeta} 
Dans l'approximation de la RPA, coefficients adimensionn\'es $\gamma$ et $\eta$ des termes $q^3$ et $q^5$ de la relation de dispersion des excitations bosoniques 
pour $q\to 0$, voir l'\'equation (\ref{eq:fleur}), en fonction de l'inverse $1/(k_F a)$ de la force des interactions.
Dans la limite CBE, ils admettent une limite finie, qui correspond \`a la relation de dispersion de Bogolioubov, voir les \'equations 
(\ref{eq:Bogo},\ref{eq:limCBE}) et les panneaux (a) et (c).
Dans la limite BCS, ils divergent~; il vaut alors mieux adimensionner le nombre d'onde $q$ par $\Delta/(\hbar c)$, ce qui revient \`a consid\'erer
les quantit\'es $\gamma \Delta^2/(m^2c^4)$ et $\eta \Delta^4/(m^4c^8)$, qui admettent une limite finie, voir les panneaux (b) et (d).
Entre ces deux limites, $\gamma$ et $\eta$ s'annulent en changeant de signe, respectivement en $1/(k_F a)\simeq-0,144$ et $1/(k_F a)\simeq-0,389$, et ils
sont respectivement faiblement positif ($\gamma\simeq 0,0838$) et faiblement n\'egatif ($\eta\simeq -0,0354$) \`a la limite unitaire.
La valeur de $\eta$ importe particulièrement au point où $\gamma$ s'annule~; en ce point, $\eta\simeq -0,0428$, donc la relation de dispersion de la RPA
est concave sur un voisinage de $q=0$.}
\end{figure}

Nous repr\'esentons sur la figure \ref{fig:gammaeta} les coefficients $\gamma$ et $\eta$ en fonction de $1/(k_F a)$, variable que nous avons préférée ici à $\Delta/\mu$.
Commentons brièvement leurs  comportements asymptotiques dans les limites CBE $1/(k_F a) \rightarrow +\infty$ et BCS $1/(k_F a) \rightarrow -\infty$ ainsi que leurs valeurs remarquables dans les régimes intermédiaires.

%%%%%%%%%%%%%%%%%%% FIGURE %%%%%%%%%%%%%%%%%%%%%%%%%
%%%%%%%%%%%%%%%%%%% ENDFIGURE %%%%%%%%%%%%%%%%%%%%%%%%%
%%%%%%%%

\subsubsection{Limite CBE}%%%%%%%% 
Dans la limite CBE, $k_F a\rightarrow 0^+$, le syst\`eme est assimilable \`a un gaz de bosons de masse $2m$ en interaction faible, de potentiel chimique 
\be
\mu_B=2\mu-E_{\rm dim},
\ee
où $E_{\rm dim}=-\hbar^2/(ma^2)$ est l'énergie interne d'un dimère \cite{Leyronas2007,Leyronas2008}. 
On sait alors que la relation de dispersion des excitations bosoniques est convexe et se r\'eduit \`a celle des excitations de Bogolioubov 
au potentiel chimique $\mu_B$ lorsque $q=o(1/a)$~\cite{CKS2006}:
\be
\hbar\omega_\qq^{\rm Bog}=\left[\frac{\hbar^2q^2}{4m}\left(\frac{\hbar^2q^2}{4m}+2\mu_B\right)\right]^{1/2},
\label{eq:Bogo}
\ee
auquel cas la vitesse du son est telle que $2mc^2 = \mu_B$. Les coefficients $\gamma$ et $\eta$ devraient donc admettre les limites suivantes~:
\be
\gamma\underset{k_Fa\to 0^+}{\to}\frac{1}{4}\ \ \ \mbox{et}\ \ \ \eta\underset{k_Fa\to 0^+}{\to}-\frac{1}{128}.
\label{eq:limCBE}
\ee
C'est bien ce que donnent les \'equations (\ref{eq:gamma}) et (\ref{eq:eta}) prises \`a la limite CBE, c'est-\`a-dire 
avec $x=O(k_F a)^{3/2}\to 0^-$ \cite{Varenna} et, comme le montre l'\'equation (\ref{eq:dDeltadmu}) apr\`es le changement de variable
$u=u'/|x|^{1/2}$,  avec $y\sim -4/x$.

%%%%%%%%
\subsubsection{Limite BCS}%%%%%%%% 
Quand $k_F a\rightarrow 0^-$, le bord inf\'erieur du continuum \`a deux excitations fermioniques (\`a nombre d'onde total $q<q_{\rm sup}$ fix\'e)
devient exponentiellement faible et 
contraint la branche d'excitation bosonique, qui ne peut p\'en\'etrer dans ce continuum, à se courber vers le bas \cite{CKS2006,Castin2015}.
Cette courbure se produit sur une \'echelle de nombre d'onde $q_c$ telle que $\hbar c q_c =\Delta$, c'est-\`a-dire telle que le terme dominant dans le développement 
\eqref{eq:fleur} soit de l'ordre de $\Delta$.
Physiquement, ceci signifie que les modes collectifs sont affectés par la structure interne des paires 
dès que leur longueur d'onde devient de l'ordre de la taille des paires dans l'espace r\'eel $\approx \hbar^2 k_F/(m \Delta)$ \cite{Strinati1998},
quantité qui est effectivement d'ordre $1/q_c$ puisque $c\propto \hbar k_F/m$ dans cette limite.

Ceci explique qualitativement pourquoi la relation de dispersion est concave aux faibles valeurs de $q$ et de $\Delta/\mu$ sur la figure \ref{fig:conc},
et donc pourquoi $\gamma <0$ dans la limite BCS.
Plus quantitativement, on s'attend \`a ce que l'énergie adimensionn\'ee $\hbar \omega_\qq/\Delta$ devienne une fonction universelle de $q/q_c=\hbar c q/\Delta$
lorsque $k_F a\to 0^-$, auquel cas tous les termes entre crochets dans l'\'equation (\ref{eq:fleur}) sont du m\^eme ordre de grandeur
en $q=q_c$, c'est-\`a-dire  $|\gamma (\hbar q_c/mc)^2|\approx 1$, $|\eta(\hbar q_c/mc)^4|\approx 1$.
En prenant la limite $x\to 0$ et $y\to 0$ dans les \'equations (\ref{eq:gamma}) et (\ref{eq:eta}), c'est bien ce que l'on obtient~:
\be
\gamma \underset{k_Fa\rightarrow 0^-}{\sim} -\frac{8}{15} \bb{\frac{mc^2}{\Delta}}^2 \ \ \mbox{et}\ \ 
\eta \underset{k_Fa\rightarrow 0^-}{\sim} \frac{136}{1575} \bb{\frac{mc^2}{\Delta}}^4,
\ee
le premier r\'esultat reproduisant celui de la r\'ef\'erence \cite{Strinati1998}.

%%%%%%%%
\subsubsection{Régimes intermédiaires}%%%%%%%% 
Dans la zone de raccordement entre CBE et BCS, $\gamma$ est une fonction croissante de $1/(k_Fa)$ et
s'annule en changeant de signe pour la valeur $x_0$ de $x=\Delta/\mu$ qui suit~:
\be
x_0\simeq 0,868567.
\ee
Ceci correspond à $1/(k_F a)\simeq-0,144292$ et confirme le résultat numérique de la figure \ref{fig:conc} ainsi que celui de la r\'ef\'erence \cite{Salasnich2015}. 
La RPA prédit donc que la limite unitaire se situe du côté où la relation de dispersion est convexe sur un voisinage de $q=0$~:
\be
\gamma\underset{(k_F a)^{-1}=0}{\simeq} 0,083769 
\ee
Le coefficient $\eta$ change lui aussi de signe, pour une valeur $x_1$ de $\Delta/\mu$ donn\'ee par
\be
x_1\simeq0,566411,
\ee
ce qui correspond à $1/(k_F a)\simeq-0,389027$. 
Il est négatif tant à la limite unitaire 
\be
\eta\underset{(k_F a)^{-1}=0}{\simeq} -0,035416,
\ee
qu'au point où $\gamma$ s'annule~:
\be
\eta(x_0) {\simeq} -0,042794.
\ee
En ce point, le signe de $\eta$ nous importe particulièrement car il détermine la concavité de la relation de dispersion au voisinage de $q=0$.
%%%%%%%%
\subsubsection{Lieu des points d'inflexion}%%%%%%%% 
Les coefficients $\gamma$ et $\eta$ permettent finalement de donner un équivalent aux faibles $q$ de la frontière entre 
les zones rouge et orange de la figure \ref{fig:conc}, autrement dit de l'ensemble des points $(q_{\rm infl}/k_\mu,\Delta/\mu)$ o\`u la dérivée seconde 
$\dd^2\omega_\qq/\dd q^2$ s'annule. 
En prenant pour $\omega_\qq$ le développement \eqref{eq:fleur} et en développant les coefficients
$\gamma(x)$ et $\eta(x)$ autour de $x=x_0$ respectivement \`a l'ordre un et \`a l'ordre z\'ero en $x-x_0$, on obtient l'équation
\be
\frac{q_{\rm infl}^2}{k_\mu^2}
\underset{x\to x_0^+}{\sim}-\frac{3\gamma'(x_0)}{10\eta(x_0)}\frac{mc^2}{\mu}(x-x_0) \simeq 2,015858(x-x_0). 
\label{eq:qinfl}
\ee
Ceci conduit à la courbe noire tiretée de la figure \ref{fig:conc}, qui rejoint l'axe $q=0$ avec une tangente horizontale.

En revanche, la fronti\`ere entre les zones rouge et orange de la figure \ref{fig:conc} rejoint le bord du domaine d'existence de la branche d'excitation collective
avec une tangente oblique car la d\'eriv\'ee troisi\`eme
de $q\mapsto \omega_\qq$ au point de contact (o\`u $q=q_{\rm sup}$) ne s'annule pas, au contraire de ce qui se passe en $q=0$.
%%%%%%%%%%%%%%%%%%%%%%%%%%%%%%%%%%%%%%%%%%%%%

\section{Conclusion}

%%%%%%%%%%%%%%%%%%%%%%%%%%%%%%%%%%%%%%%%%%%%%
Pour un gaz spatialement homog\`ene de fermions de spin $1/2$ non polaris\'e et \`a temp\'erature nulle, nous avons obtenu l'énergie propre $\hbar \omega_\qq$
de la branche d'excitation bosonique pr\'edite par la RPA jusqu'à l'ordre $5$ en le nombre d'onde $q$ au voisinage de $q=0$,
avec des coefficients qui sont des fractions rationnelles en 
$\Delta/\mu$ et $(\partial\Delta/\partial\mu)_a$, la seconde variable pouvant être analytiquement reliée à la première {\sl via} l'équation d'état de la th\'eorie BCS. 
Ceci nous permet de montrer analytiquement que la relation de dispersion $q\mapsto \omega_\qq$ est concave sur un voisinage de $q=0$ lorsque $1/(k_F a)$ est compris entre $-\infty$ et une valeur proche de $-0,144$, 
point où la première correction à l'énergie phononique est quintique et faiblement négative. 
Pour $-0,144<1/(k_F a)<0,157$ la branche est convexe sur voisinage de $q=0$ mais redevient concave lorsque $q$ augmente (et le demeure jusqu'\`a la valeur maximale
accessible de $q$ si $1/(k_F a)<0,022$, sinon elle redevient convexe). Au-delà de $1/(k_F a)=0,157$, elle est entièrement convexe sur son domaine d'existence. 

Une application directe de notre développement du spectre (\ref{eq:fleur})  est de déterminer si une excitation collective de faible vecteur d'onde $\qq$ peut ou non décroître 
par un processus de Beliaev, c'est-à-dire en émettant deux excitations collectives de vecteurs d'onde
 $\qq_1$ et $\qq_2=\qq-\qq_1$. 
Ce processus est autorisé par la loi de conservation de l'énergie à condition que $\omega_\qq > \Omega_\qq^{\rm inf}$
où $\Omega_\qq^{\rm inf}=\inf_{\qq_1} (\omega_{\qq_1}+\omega_{\qq-\qq_1})$ est le bord inférieur du continuum à deux excitations de vecteur d'onde total $\qq$ fixé.
Lorsque $\Delta/\mu$ est loin de la valeur critique $x_0\simeq 0.869$ où le coefficient $\gamma$ de $q^3$ s'annule,
la relation de dispersion est soit entièrement convexe soit entièrement concave aux faibles $q$, et l'amortissement de Beliaev est respectivement permis ou interdit.
En revanche, si $\Delta/\mu$ est proche de la valeur critique $x_0$, la relation de dispersion possède un point d'inflexion aux faibles $q$.
Dans ce cas, nous appliquons à l'équation (\ref{eq:fleur}) l'analyse fournie par la référence \cite{Castin2015} 
et nous trouvons $\Omega_\qq^{\rm inf}=\min(2\omega_{\qq/2},\omega_\qq)$
\footnote{Puisque la relation de dispersion $q\mapsto\omega_\qq$ est ici une fonction croissante, seuls les cas (i) et (ii) 
de l'équation (12) de la référence \cite{Castin2015}, où $\qq_1$ et $\qq-\qq_1$ sont de même direction et de même sens
au minimum de $\omega_{\qq_1}+\omega_{\qq-\qq_1}$, sont à prendre en compte.
Il ne reste qu'à chercher le minimum de la fonction $f(q_1)=\omega(q_1)+\omega(q-q_1)$ sur l'intervalle $[0,q/2]$,
où $\omega(q)=\omega_\qq$  est donné par l'équation (\ref{eq:fleur}).
En utilisant le paramétrage $q_1=(1-t)q/2$, avec $t\in [0,1]$, nous trouvons que $f(q_1)$ est une fonction parabolique concave de $t^2$.
Son minimum est donc atteint sur l'un des bords de l'intervalle et vaut soit $f(0)=\omega_\qq$ soit $f(q/2)=2\omega_{\qq/2}$.}.
L'amortissement de Beliaev est donc permis à condition que
\be
\left(\frac{\hbar q}{mc}\right)^2 < \frac{8\gamma}{5|\eta|}
\ee
à l'ordre dominant en $\gamma$, c'est-à-dire en $\Delta/\mu-x_0$.
À température basse mais non nulle, d'autres mécanismes d'amortissement sont possibles~:
(i) le mécanisme de Landau $\qq+\qq_1\to \qq_2$ est interdit dès lors que celui de Beliaev l'est puisqu'il peut être vu comme l'inverse d'un processus de Beliaev
avec un vecteur d'onde $\qq_2$ de module $>q$;
(ii) le processus d'ordre supérieur de Landau-Khalatnikov $\qq+\qq_1\to \qq_2+\qq_3$ est lui toujours permis, mais il est sous-dominant dès lors que les processus de Beliaev ou Landau sont présents.

Nos résultats sur la concavité au voisinage de $q=0$ peuvent être testés expérimentalement dans un gaz d'atomes fermioniques froids 
pi\'eg\'e dans un potentiel \`a fond plat \cite{Hadzibabic2013} en 
mesurant des effets dissipatifs, comme le taux d'amortissement des excitations collectives à basse température, des effets dispersifs, comme l'\'etalement
au cours de sa propagation dans le gaz d'un paquet d'ondes sonores cr\'e\'e par une impulsion lumineuse \cite{Ketterle1997,Grimm2013}, ou en acc\'edant
directement \`a la relation de dispersion \`a basse temp\'erature {\sl via} le facteur de structure dynamique du gaz de fermions par excitation de Bragg \`a un vecteur d'onde $\qq$ bien d\'efini 
\cite{Ketterle1999,Davidson2002,Hannaford2010}
\footnote{
Dans la référence \cite{Davidson2002} le spectre est mesuré avec une incertitude de $\pm 18\,$Hz aux faibles $q$. Pour une température de Fermi typique
 $T_F=1\,\mu$K, ceci correspond à une incertitude de $\pm 2\times 10^{-3}$ sur $\hbar\omega_\qq/\mu$, c'est-à-dire sur l'axe des ordonnées de la figure \ref{fig:disp}. Ceci indique
que la convexité de $q\mapsto \omega_\qq$ peut être déterminée expérimentalement dans un gaz unitaire de fermions.}

Coupl\'es \`a des \'equations cin\'etiques sur les nombres d'occupation des modes collectifs,
nos r\'esultats ouvrent aussi la voie à une d\'etermination analytique du coefficient de diffusion de phase \`a basse temp\'erature, 
donc de la limite intrins\`eque fondamentale au temps de cohérence du condensat de paires dans un gaz de fermions de taille finie \cite{KCS2015f}.

%%%%%%%%%%%%%%%%%%%%%%%%%%%%%%%%%%%%%%%%%%%%%

%%%%%%%%%%%%%%%%%%%%%%%%%%%%%%%%%%%%%%%%%%%%%
\appendix
%%%%%%%%%%%%%%%%%%%%%%%%%%%%%%%%%%%%%%%%%%%%%

\section{Expression d'intégrales en termes des variables $x$ et $y$}
\label{app:integrales}
%%%%%%%%%%%%%%%%%%%%%%%%%%%%%%%%%%%%%%%%%%%%%
Dans le développement de $I_{++}(\omega_\qq,q)$, $I_{--}(\omega_\qq,q)$ et $I_{+-}(\omega_\qq,q)$ à faible $q$, et apr\`es adimensionnement des vecteurs d'onde 
par $k_\Delta$ comme dans l'\'equation \eqref{eq:dDeltadmu}, 
\`a savoir $k=u k_\Delta$, nous rencontrons des intégrales de la forme
\bea
I_{n,p} &=& \frac{k_\Delta^3}{2\pi^2\rho}\int_0^{+\infty}\dd u\frac{u^{2p+2}}{\epsilon_u^n}\label{eq:notationunifiee1} \\
J_{n,p} &=& \frac{k_\Delta^3}{2\pi^2\rho}\int_0^{+\infty}\dd u\frac{u^{2p+2}\xi_u}{\epsilon_u^n} \label{eq:notationunifiee2}
\eea
avec $n\in2\mathbb{N}^*+1$, $p\in\mathbb{N}$, 
\bea
\xi_u     &=& u^2-\frac{1}{x} \\
\epsilon_u&=& \sqrt{\xi_u^2+1}
\eea
et la densit\'e totale $\rho$ du gaz est donn\'ee par l'\'equation (\ref{eq:rho}).
L'int\'egrale donnant $I_{n,p}$ est convergente pour $n-p\geq2$, tandis que celle donnant $J_{n,p}$
l'est pour $n-p\geq 3$. 
Les intégrales que nous rencontrons dont l'intégrande n'est pas fonction uniquement du module de $\kk$ se ramènent facilement à \eqref{eq:notationunifiee1} ou \eqref{eq:notationunifiee2} 
par int\'egration angulaire~:
\begin{multline}
\int \dd^3k f(k) \bb{\frac{\hbar^2\kk\cdot\qq}{m}}^{2p} = \\ 
\frac{4\pi}{1+2p} \bb{\frac{\hbar^2q^2}{m}}^{p} \int_0^{+\infty} \dd k k^2 f(k) \bb{\frac{\hbar^2 k^2}{m}}^{p} 
\end{multline}
o\`u $f(k)$ est une fonction arbitraire du module de $\kk$.

\'Etablissons d'abord les quatre relations de r\'ecurrence suivantes~:
\bea
I_{n,p} &=& \frac{n-3}{n-2}I_{n-2,p}-\frac{2p+1}{2(n-2)} J_{n-2,p-1} \label{eq:Inp}\\
J_{n,p} &=& \frac{2p+1}{2(n-2)} I_{n-2,p-1} \label{eq:Jnp}\\
I_{n,p} &=& J_{n,p-1}+\frac{I_{n,p-1}}{x} \label{eq:Inpbis}\\
J_{n,p} &=&\frac{J_{n,p-1}}{x} +  I_{n-2,p-1}- I_{n,p-1}\label{eq:Jnpbis},
\eea
avec la condition $1\leq p\leq n-4$ dans la premi\`ere relation,  $1\leq p\leq n-3$ dans la deuxi\`eme, $1\leq p\leq n-2$  dans la troisi\`eme 
et $1\leq p\leq n-3$ dans la derni\`ere.
Pour établir la relation \eqref{eq:Inp}, on int\`egre par parties l'intégrale 
\be
\frac{k_\Delta^3}{2\pi^2\rho}\int_0^{+\infty}\dd u\frac{u^{2p+2}\xi_u^2}{\epsilon_u^n}=I_{n-2,p}-I_{n,p},
\ee
en d\'erivant la fonction $u\mapsto u^{2p+1} \xi_u$. Pour établir la relation \eqref{eq:Jnp}, on int\`egre par parties l'int\'egrale d\'efinissant
$J_{n,p}$ dans l'\'equation (\ref{eq:notationunifiee2}), en d\'erivant la fonction $u\mapsto u^{2p+1}$.
Dans les deux cas, on note que la fonction $u\mapsto u \xi_u/\epsilon_u^n$ admet comme primitive $u\mapsto -[2(n-2)\epsilon_u^{n-2}]^{-1}$.
Enfin, on \'ecrit simplement $u^{2p+2}=u^{2p}(\xi_u+x^{-1})$ dans l'int\'egrande de (\ref{eq:notationunifiee1}) pour obtenir (\ref{eq:Inpbis}),
et $u^{2p+2}\xi_u=u^{2p}(\epsilon_u^2-1+x^{-1}\xi_u)$ dans l'int\'egrande de (\ref{eq:notationunifiee2}) pour obtenir (\ref{eq:Jnpbis}).
Cette proc\'edure g\'en\'eralise celle de la r\'ef\'erence \cite{Strinati1998}.

Montrons ensuite par r\'ecurrence que les relations (\ref{eq:Inp},\ref{eq:Jnp},\ref{eq:Inpbis},\ref{eq:Jnpbis})
permettent d'exprimer $I_{n,p}$ et $J_{n,p}$ en fonction de $I_{3,0}$ et $J_{3,0}$,
pour tout $n$ impair $\geq 3$ et pour tout $p$ positif dans le domaine d'existence des int\'egrales.
Soit un tel $n$.  Supposons que l'on connaisse tous les $I_{n,p}$, $0\leq p\leq n-2$, et tous les $J_{n,p}$, $0\leq p\leq n-3$.
Alors $(i)$ $I_{n+2,1}$ et $J_{n+2,1}$ s'en d\'eduisent gr\^ace \`a (\ref{eq:Inp}) et (\ref{eq:Jnp}), $(ii)$ on obtient gr\^ace \`a (\ref{eq:Jnpbis})
et (\ref{eq:Inpbis}) le syst\`eme de Cramer sur $I_{n+2,0}$ et $J_{n+2,0}$~:
\bea
x^{-1} J_{n+2,0} - I_{n+2,0} &=& J_{n+2,1} - I_{n,0} \\
J_{n+2,0} +x^{-1} I_{n+2,0} &=& I_{n+2,1}
\eea
que l'on r\'esout, $(iii)$ on utilise (\ref{eq:Inpbis}) et (\ref{eq:Jnpbis}) pour acc\'eder aux valeurs de $I_{n+2,p}$ et $J_{n+2,p}$ pour $p\geq 2$.
On initialise la r\'ecurrence \`a $n=3$, en exprimant $I_{3,1}$ en fonction de $I_{3,0}$ et $J_{3,0}$ gr\^ace \`a (\ref{eq:Inpbis}).

On relie enfin $I_{3,0}$ et $J_{3,0}$ à $x$ et $y$ grâce aux relations \eqref{eq:dDeltadmu} et \eqref{eq:thetasurN} données dans le corps de l'article, 
qui s'\'ecrivent avec les notations de cette annexe sous la forme $y=J_{3,0}/I_{3,0}$ et $I_{3,0}=3x/[2(1+xy)]$.

%\bibliography{biblio}
%\bibliographystyle{unsrt-fr}

\providecommand*\hyphen{-}